\documentclass[lettersize,journal]{IEEEtran}

 \usepackage{array} 
\usepackage{textcomp}
\usepackage{stfloats}
\usepackage{url}
\usepackage{verbatim}
\usepackage{graphicx} 
\usepackage{subfigure} 
\usepackage{cite}
\usepackage{amsmath,graphicx}
\usepackage{amssymb}
\usepackage{algpseudocode}
\usepackage{amsfonts} 
\usepackage{fancyhdr}

\usepackage{cases}
\usepackage{extarrows}
\usepackage{algorithm}
\usepackage{multirow,tabularx}
\usepackage{mathtools}
\usepackage{xcolor}
\usepackage[english]{babel}
\usepackage{caption}
\usepackage{autobreak}
\usepackage{bbm}

\usepackage[colorlinks,linkcolor=red]{hyperref}

\begin{document}
\title{ Electromagnetic Channel Modeling and Capacity Analysis for HMIMO Communications}
\author{\IEEEauthorblockN{Li Wei, Shuai S. A. Yuan, Chongwen Huang, Jianhua Zhang,~\IEEEmembership{Senior Member,~IEEE},  Faouzi Bader, ~\IEEEmembership{Senior Member,~IEEE},  
Zhaoyang Zhang, Sami Muhaidat, ~\IEEEmembership{Senior Member,~IEEE}, M\'{e}rouane Debbah,~\IEEEmembership{Fellow,~IEEE}  and Chau Yuen,~\IEEEmembership{Fellow,~IEEE}} 
\thanks{ L. Wei and C. Yuen are with School of Electrical and Electronics Engineering, Nanyang
Technological University, Singapore 639798 (e-mails: \{l\_wei, chau.yuen\}@ntu.edu.sg).}
\thanks{ S. S. A. Yuan, C. Huang and Z. Zhang  are with College of Information Science
and Electronic Engineering, Zhejiang University, Hangzhou 310027, China  (e-mails:
\{shuaiyuan1997, chongwenhuang, ning\_ming\}@zju.edu.cn).
 }
 \thanks{J. Zhang is with the State Key Laboratory of Networking and Switching Technology, Beijing University of Posts and Telecommunications, Beijing 100876, China (e-mail: jhzhang@bupt.edu.cn).}

\thanks{F. Bader is with the Technology Innovation Institute, Masdar City, Abu Dhabi, United
Arab Emirates (e-mail: carlos-faouzi.bader@tii.ae).}

\thanks{S. Muhaidat is with the KU 6G Research Center, Department
of Electrical Engineering and Computer Science, Khalifa University, Abu Dhabi
127788, UAE, and also with the Department of Systems and Computer Engineering, Carleton University, Ottawa, ON K1S 5B6, Canada (e-mail: muhaidat@ieee.org).}

 \thanks{M. Debbah is with KU 6G Research Center, Khalifa University of Science and Technology, P O Box 127788, Abu Dhabi, UAE (email: merouane.debbah@ku.ac.ae).}

 \thanks{The code is available at \url{https://github.com/WEILI-NTU/Stochastic-EM-Channel-Modeling}.}
 
}

\maketitle

\thispagestyle{fancy}

\begin{abstract}
Advancements in emerging technologies, e.g., reconfigurable intelligent surfaces and holographic MIMO (HMIMO), facilitate unprecedented manipulation of electromagnetic (EM) waves, significantly enhancing the performance of wireless communication systems. To accurately characterize the achievable performance limits of these systems, it is crucial to develop a universal EM-compliant channel model. This paper addresses this necessity by proposing a comprehensive EM channel model tailored for realistic multi-path environments, accounting for the combined effects of antenna array configurations and propagation conditions in HMIMO communications. Both polarization phenomena and spatial correlation are incorporated into this probabilistic channel model. Additionally, physical constraints of antenna configurations, such as mutual coupling effects and energy consumption, are integrated into the channel modeling framework.  Simulation results validate the effectiveness of the proposed probabilistic channel model, indicating that traditional Rician and Rayleigh fading models cannot accurately depict the channel characteristics and underestimate the channel capacity. More importantly, the proposed channel model outperforms free-space Green's functions in accurately depicting both near-field gain and multi-path effects in radiative near-field regions. These gains are much more evident in tri-polarized systems, highlighting the necessity of polarization interference elimination techniques. Moreover, the theoretical analysis accurately verifies that capacity decreases with expanding communication regions of two-user communications.
 
\end{abstract}


\section{Introduction} \label{sec:intro}
To support the explosive growth of data and the proliferation of devices in wireless communications, increasing the number of antennas is an effective method, yielding the massive multiple-input multiple-output (MIMO) systems \cite{5595728,shlezinger2020dynamic_all,9475156,10505154,10542433,JINDAN_CS}, such as extremely large antenna arrays (ELAA) \cite{ramezani2023near} and holographic MIMO (HMIMO) systems \cite{9136592}. In ELAA, the antennas are distributed over a large geographical area and the antenna separation is on the order of wavelength $\lambda$, resulting in a large array aperture. Benefiting from the large number of antenna elements and array aperture, the ELAA could provide a high area throughput, where the users are located in both the near-field and far-field regions. HMIMO systems  \cite{RISE6G_COMMAG,9779586,9136592, 8741198,10663346,10130641} integrate a large number of electrically small antenna elements in a compact arrangement to create a nearly continuous aperture, thus achieving high spatial resolution and unprecedented electromagnetic (EM) wave manipulation, where the superdirectivity is achievable. Benefiting from these merits, both techniques are capable of improving throughput and spectral efficiency \cite{bjornson2019massive,10232975,9136592}, and can be applied in many scenarios, such as wireless power transfer and positioning \cite{9136592,9530717}.   

Compared to traditional MIMO systems, these emerging wireless communication techniques exhibit distinct characteristics that significantly enhance performance. The primary difference lies in the shift from conventional far-field regions to near-field zones, where evanescent EM waves can be exploited. The evanescent EM waves carry higher-order powers, yielding additional performance gain through near-field effects. As the distance between the transmitter and receiver increases, the evanescent waves vanish, leaving only propagating waves. Consequently, the capacity of far-field communications is lower than that of near-field communications. Specifically, tri-polarization (TP) can be achieved in the near-field regime, while dual-polarization (DP) can be obtained in the far-field zone due to the vanishing polarization state \cite{FranceschettiCapacityWirelessNetworks2009, FranceschettiInformationCarriedScattered2015, 9495942,DENGZhiji41}. On the other hand, the second difference is the mutual coupling effects induced by small spacing ($<\lambda/2$) between adjacent antennas, as observed in HMIMO systems. Such a design enables fine spatial resolution, making it possible to achieve superdirective beams with low power consumption. The dense antenna configuration with spacing ($\ll \lambda/2$) can manipulate EM waves at a high level, achieving superdirectivity at a low cost.     

\begin{figure}  
	\begin{center}
		{\includegraphics[width=0.45 \textwidth]{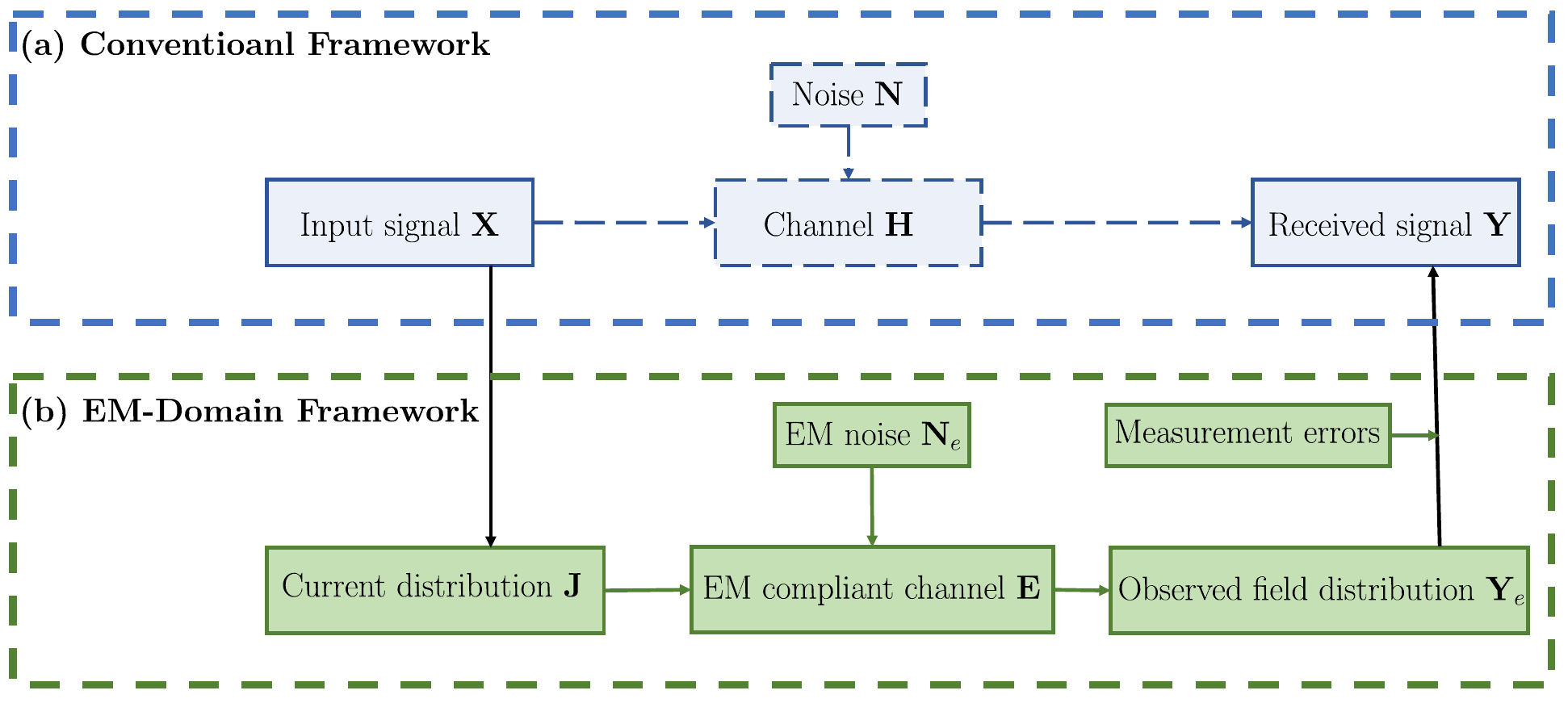}}  
		\caption{The framework of wireless communications. (a) Conventional framework; (b) EM-domain framework. }  
		\label{fig:InputOutput} 
	\end{center}
\end{figure}

Therefore, EM-domain signal processing introduces new challenges to traditional communication systems by exploiting the inherent EM propagation. As shown in Fig.~\ref{fig:InputOutput}, the traditional input signal $\mathbf{X}$ is modulated by current distribution $\mathbf{J}$ that excites distinct EM waves, thus, the current $\mathbf{J}$ is equivalent to the transmitted signal but in a more fundamental way.  Similarly, the field $\mathbf{Y}_e$ observed by the receiver is processed to obtain the received signal $\mathbf{Y}$, implying potential measurement error. Additionally, the channel matrix $\mathbf{H}$ maps the input signal to the received signal, while the EM-compliant channel $\mathbf{E}$ is the mapping between the current distribution and the observed field. Unlike the conventional framework that ignores the physical limitations, such as the antenna configurations (e.g., antenna size, spacing, etc.), the EM-domain framework investigates the underlying signal transmission behavior, considering the impacts of both physical settings and complex near-field effects on the overall system performance.

Nevertheless, traditional channel models fail to depict such a complicated EM transmission process. First, near-field communications are characterized by the spherical wavefront, distinct from the planar wavefront in far-field communications, making traditional far-field channel models inapplicable. In addition, conventional channel models simplify the transmission process using certain approximations, which underestimate the capacity of emerging communications in both near-field and far-field regions. At the same time, while the antenna array with spacing ($<\lambda/2$) generates the mutual coupling effects and fine spatial resolution, the energy efficiency is reduced and the radiation pattern is deformed. The traditional independent and identically distributed (i.i.d.) Gaussian channel models assume no coupling effects with half-wavelength spacing, rendering them incapable of depicting the inherent mutual coupling and overestimating the performance gain \cite{9497725,10443720}. It has been proved that the capacity or degrees of freedom cannot consistently increase with more antenna elements, and the performance gains are typically constrained by the array size \cite{4447351, 7805306, 9650519}. Furthermore, the physical antenna configuration impacts the quality factor, which is inversely proportional to bandwidth and influences system performance \cite{8024037, 1417209,9399839,8998551,10628002}. Therefore, to describe the joint impacts of antenna array configuration and the communication zones, it is crucial to investigate the underlying electromagnetic EM waves that carry the information content. Based on the fundamentals of EM transmission and antenna array design, several EM-compliant channel models have been proposed.

There are primarily two types of channel models: deterministic models and stochastic models. In deterministic modeling, the scalar Green’s function and the dyadic Green’s function, which depict the point-to-point EM wave propagation between the field and a point source, can be utilized for line-of-sight (LoS) channel modeling in communication systems. For instance, to accurately describe polarization and spatial correlation effects, a dyadic Green’s function-based channel model for HMIMO systems has been proposed in \cite{10103817}, employing mathematical approximation and integral computation over small antenna elements. However, this approach is primarily applicable to LoS scenarios and incurs excessively high computational costs when dealing with random scatterers, making it ineffective at accurately capturing multi-path effects caused by these scatterers. To characterize such random scattered environments, the work in \cite{9724113} proposed a Fourier plane-wave expansion based model, which discretizes the EM fields in the wavenumber domain and incorporates randomness in variances of expansion coefficients.  Nevertheless, this model is best suited for regular scatterers and may be less effective or straightforward when describing irregular scatterers, which are commonly encountered in wireless communication environments. To characterize multi-path effects caused by random scatterers, a stochastic Green's function is proposed in \cite{10018012} to depict polarization effects and spatial correlation in the multi-path environment. This model is both tractable and statistical, allowing for effective analysis of wave physics in complex wave-chaotic environments \cite{10381502}. Specifically, as numerous planar waves are transmitted or reflected between the source point and the observation field, the distribution of the eigenfunctions can be approximated by the superposition of these planar waves, leading to the derivation of the statistical wave model.
 
Resorting to the stochastic Green's function in \cite{10018012}, we presented a stochastic channel model in HMIMO systems, which accounts for the multi-path environment along with physical constraints. Notably, the proposed channel model can be easily extended to general communication systems. Specifically, the LoS channel is modeled using dyadic Green's function, while the non-line-of-sight (NLoS) channel is modeled using the eigenfunction expansion, where the means and variances of co-polarized and cross-polarized components are derived. Additionally, the model incorporates the $K$-factor and mutual coupling effects in HMIMO systems. Based on this probabilistic channel model, we investigate the theoretical capacity bound and capacity region for two-user scenarios. The contributions of this paper can be summarized as follows:  
	\begin{itemize}
		\item Building on the stochastic dyadic Green's function, we present a universal channel model for HMIMO systems in the multi-path environment, which captures both polarization characteristics and spatial correlation effects. Specifically, the distribution of polarization components is derived to model NLoS channels, and the coupling effects are also embedded in the proposed channel model. The physical implications of the proposed channel model, including the radiation intensity, radiation power, and antenna gain, are also discussed.
		\item We further evaluate the capacity of HMIMO systems within the EM-domain framework. Specifically,  the lower and upper bounds of the single-input-single-output (SISO) systems and multiple-input-single-output (MISO) systems are derived in terms of transceiver distance. Additionally, the approximation of capacity region for two-user cases is investigated in different communications regions.   
		\item  Our simulation results demonstrate the effectiveness of the proposed channel through comparison with baselines. It is validated that conventional channel models (Rician model and i.i.d. Rayleigh model) are incapable of depicting channel correlations and capacity variations over distance. Additionally, the proposed channel model describes polarization effects and spatial correlation in both near-field and far-field communications. It demonstrates that additional capacity gains are achieved through near-field effects in near-field communications and multi-path effects in far-field communications, alongside polarization gains. The theoretical analysis accurately predicts the practical capacity performance across all communication regions as well.  
	\end{itemize}       

The remainder of this paper is organized as follows. In Section II, the stochastic Green's function based channel modeling for HMIMO systems is introduced.  Section III derives the theoretical capacity bound of SISO/MISO systems and the capacity region for two-user cases. Section IV presents numerical simulations. Finally, the paper’s conclusions are drawn in Section V.

\textit{Notation}: Fonts $a$, $\mathbf{a}$, and $\mathbf{A}$ represent scalars, vectors, and matrices, respectively. $\mathbf{A}^T$, $\mathbf{A}^{\dagger}$, $\mathbf{A}^{-1}$, and $\|\mathbf{A}\|_f$ denote transpose, Hermitian (conjugate transpose), inverse (pseudo-inverse), and Frobenius norm of $ \mathbf{A} $, respectively. $\mathbf{A}_{i,j}$ or $[\mathbf{A}]_{i,j}$ represents $\mathbf{A}$'s $(i,j)$th element.   $\mathbf{I}_n$ (with $n\geq2$) is the $n\times n$ identity matrix. $\otimes$ denotes an outer product. $\operatorname{Re}[\cdot]$ is the real part of the argument.   Finally, notation $\mathbb{E}[\cdot]$ and $\mathbb{V}[\cdot]$ denote the expectation and variance of the argument, respectively. $\hat{\mathbf{x}}, \hat{\mathbf{y}}$, and $\hat{\mathbf{z}}$ denote axes in Cartesian coordinate system.

\section{Channel Modeling} \label{sec:ChannelModel}

In this section, we first review the stochastic Green's function proposed by S. Lin  \cite{10018012, 8952896}, a statistical wave model to describe multipath effects in the electronic enclosure. We then derive the channel modeling based on stochastic Green's functions. In addition, we incorporate the mutual coupling effects to establish a more practical and probabilistic channel model. 

\subsection{Preliminaries of Stochastic Green's Function}

The dyadic Green's function can be expanded using eigenfunction expansion given by 
\begin{equation} \label{equ:Green's Function}
\overline{\overline{\mathbf{G}}}\left(\mathbf{r}, \mathbf{r}^{\prime}\right)=\sum_i \frac{\boldsymbol{\Psi}_i\left(\mathbf{r}, k_i\right) \otimes \boldsymbol{\Psi}_i\left(\mathbf{r}^{\prime}, k_i\right)}{k^2-k_i^2-j \frac{k^2}{\tilde{Q}}},
\end{equation}
where $\boldsymbol{\Psi}_i$ and $k_i$ are $i$-th eigenfunction and eigenvalue, respectively.  $\mathbf{r}$ and $\mathbf{r}'$ are source point and receiving point, respectively. $k=\frac{2\pi}{\lambda}$ is the wavenumber and  $\tilde{Q}$ is the cavity quality factor that is dependent on transmission environment settings.   

The vector components of eigenfunction at the transmitter and receiver are expressed by correlated Gaussian random variables ${\Psi}_i^x(\mathbf{r}), {\Psi}_i^y(\mathbf{r})$ and ${\Psi}_i^z(\mathbf{r})$, i.e.,  \cite{10018012}
\begin{equation}
\begin{aligned}
\boldsymbol{\Psi}_i\left(\mathbf{r}, k_i\right) & \simeq {\Psi}_i^x(\mathbf{r}) \hat{\mathbf{x}}+{\Psi}_i^y(\mathbf{r}) \hat{\mathbf{y}}+{\Psi}_i^z(\mathbf{r}) \hat{\mathbf{z}}, \\
\boldsymbol{\Psi}_i\left(\mathbf{r}^{\prime}, k_i\right) & \simeq {\Psi}_i^x(\mathbf{r}') \hat{\mathbf{x}}+ {\Psi}_i^y(\mathbf{r}') \hat{\mathbf{y}}+{\Psi}_i^z(\mathbf{r}')\hat{\mathbf{z}},
\end{aligned}
\end{equation}
where
\begin{equation}
    \begin{aligned}
  {\Psi}_i^x(\mathbf{r}) \simeq & \lim _{N \rightarrow \infty} \sum_{n=1}^N\left[a _ { n } \left(-\cos \psi_n \sin \phi_n-\right.\right. \\
& \left.\left.\sin \psi_n \cos \phi_n \cos \theta_n\right) \cos \left(k_i \hat{\mathbf{e}}_n \cdot \mathbf{r}+\beta_n\right)\right],\\
{\Psi}_i^y(\mathbf{r}) \simeq & \lim _{N \rightarrow \infty} \sum_{n=1}^N\left[a _ { n } \left(\cos \psi_n \cos \phi_n-\right.\right. \\
& \left.\left.\sin \psi_n \sin \phi_n \cos \theta_n\right) \cos \left(k_i \hat{\mathbf{e}}_n \cdot \mathbf{r}+\beta_n\right)\right] \\
{\Psi}_i^z(\mathbf{r}) \simeq & \lim _{N \rightarrow \infty} \sum_{n=1}^N\left[a_n \sin \psi_n \sin \theta_n \cos \left(k_i \hat{\mathbf{e}}_n \cdot \mathbf{r}+\beta_n\right)\right],
\end{aligned}
\end{equation}
with $a_n$ being the $n$-th plane wave's amplitude that satisfies $\langle a_m a_n\rangle=\frac{2}{NV}\delta_{mn}$, in which $V$ is the volume of the cavity.   $\psi_n$, $\phi_n$, and $\theta_n$ denote the polarization angle, azimuth angle, and elevation angle of the $n$-th plane wave, respectively. Both $\psi_n$ and $\phi_n$ are assumed to be uniformly distributed over $[0, 2\pi]$, while $\theta_n$ is assumed to be uniformly distributed over $[0, \pi]$.  $\hat{\mathbf{e}}_n$ is the direction of the $n$-th plane wave.  $\{(w_i^x)',(w_i^y)',(w_i^z)'\}$ is similar to the above equations by replacing $\mathbf{r}$ with $\mathbf{r}'$.

Accordingly, the outer product in the numerator in \eqref{equ:Green's Function} is written as 
\begin{equation}
\begin{aligned}
&\overline{\overline{\mathbf{D}}}\left(\mathbf{r}, \!\mathbf{r}^{\prime} ; k_i\right)\!\!=\!\!  {\Psi}_i^x(\mathbf{r}) {\Psi}_i^x(\mathbf{r}') \hat{\mathbf{x}} \hat{\mathbf{x}}+{\Psi}_i^x(\mathbf{r}) {\Psi}_i^y(\mathbf{r}')  \hat{\mathbf{x}} \hat{\mathbf{y}}+{\Psi}_i^x(\mathbf{r}) {\Psi}_i^z(\mathbf{r}')  \hat{\mathbf{x}} \hat{\mathbf{z}} \\
&\qquad  +{\Psi}_i^y(\mathbf{r}) {\Psi}_i^x(\mathbf{r}') \hat{\mathbf{y}} \hat{\mathbf{x}}+{\Psi}_i^y(\mathbf{r}) {\Psi}_i^y(\mathbf{r}')  \hat{\mathbf{y}} \hat{\mathbf{y}}+{\Psi}_i^y(\mathbf{r}) {\Psi}_i^z(\mathbf{r}') 
 \hat{\mathbf{y}} \hat{\mathbf{z}} \\
&\qquad +{\Psi}_i^z(\mathbf{r}) {\Psi}_i^x(\mathbf{r}') \hat{\mathbf{z}} \hat{\mathbf{x}}+{\Psi}_i^z(\mathbf{r}) {\Psi}_i^y(\mathbf{r}')  \hat{\mathbf{z}} \hat{\mathbf{y}}+{\Psi}_i^z(\mathbf{r}) {\Psi}_i^z(\mathbf{r}')  \hat{\mathbf{z}} \hat{\mathbf{z}}.
\end{aligned}
\end{equation}

The stochastic Green's function is given by \cite{10018012} 
\begin{equation}
\overline{\overline{\mathbf{G}}}_{\mathrm{S}}\left(\mathbf{r}, \!\mathbf{r}^{\prime} ; k\right)\!=\!\operatorname{Re}\left[\overline{\overline{\mathbf{G}}}_0\left(\mathbf{r}, \!\mathbf{r}^{\prime} ; k\right)\right]\!+\!\!\sum_m \frac{\overline{\overline{\mathbf{D}}}\left(\mathbf{r},\! \mathbf{r}^{\prime} ; k_m\right)}{\tilde{\lambda}_m-j \alpha} \frac{k V}{2 \pi^2},
\end{equation}
where the normalized spacing $\tilde{\lambda}_m$ between adjacent eigenvalues and antenna-dependent parameter $\alpha$ are given by \eqref{equ:lambda}, and the dyadic Green's function $\overline{\overline{\mathbf{G}}}_0$ is given in \eqref{equ:dyadicGreen}. 

The above derivation offers novel insights into developing a multi-path channel model for HMIMO systems, especially the full-polarization channel model applicable to both far-field and near-field communications. This model is further detailed in the next section.

\subsection{Channel Modeling of Isolated Antenna in HMIMO systems}
 \begin{figure}  
	\begin{center}
		{\includegraphics[width=0.5 \textwidth]{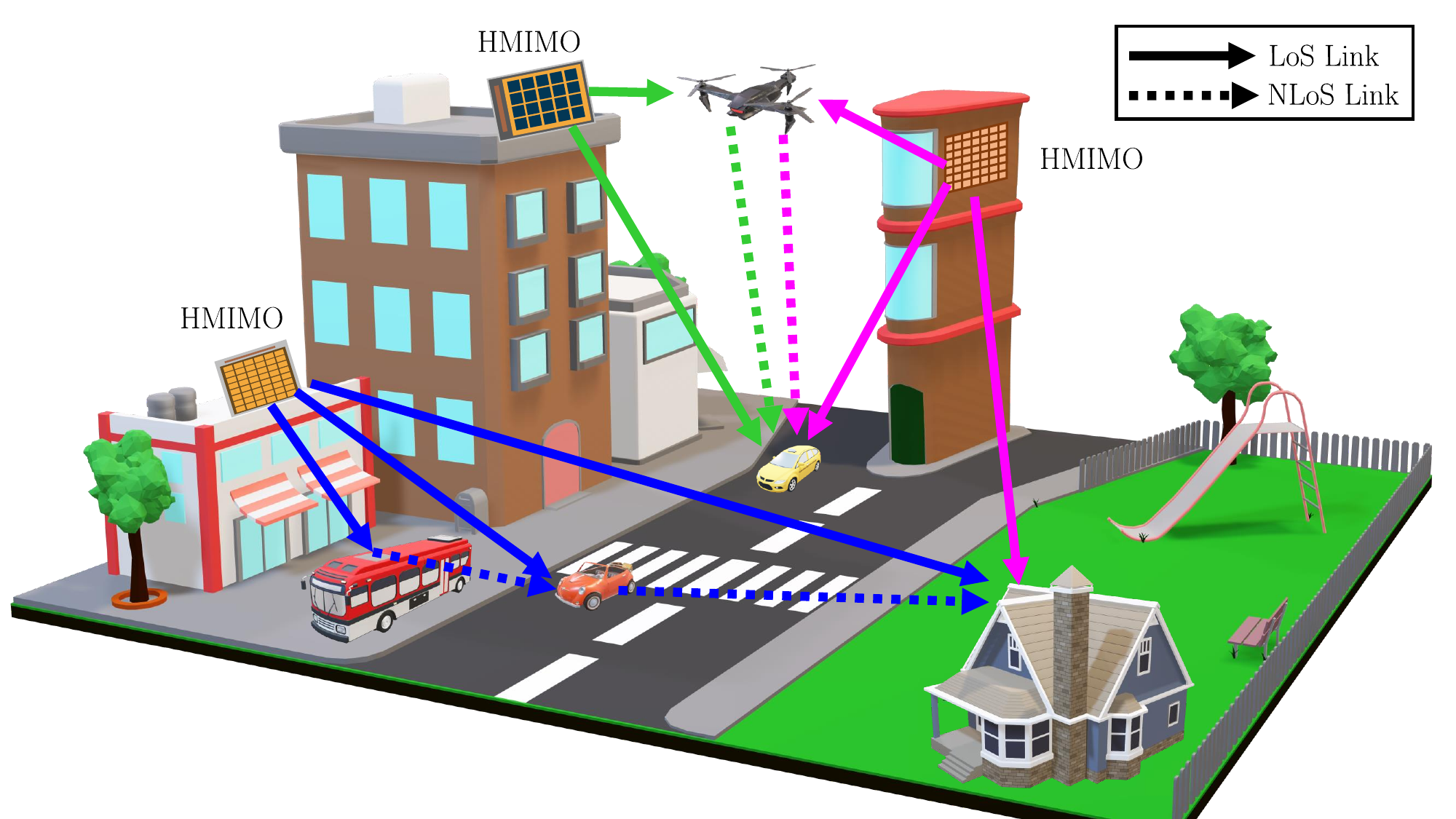}}  
		\caption{Illustration of HMIMO systems in the scattered environment.}  
		\label{fig:SystemModel} 
	\end{center}
\end{figure}
 
Consider a transmitter HMIMO system consisting of $N_s$ rectangular patch antennas, each with lengths $s_x$ and $s_y$ in $\hat{\mathbf{x}}$ and $\hat{\mathbf{y}}$ directions, as shown in Fig.~\ref{fig:SystemModel}. The spacing between two adjacent antennas is denoted as $d_s$. The peak current at the center is denoted by $I$. At the receiver end, each user is equipped with $N_r$ rectangular patch antennas. The distance between the transmitter and receiver  is $R_{mn}, m=1,\ldots, N_r, n=1,\ldots,N_s$. We first investigate the transmission between the individual transmitter antenna and receive antenna, and omit the subscript of $R_{mn}$ for simplicity. 

Given the presence of random scatterers in practical environments, the paths experienced by the transmitted signal can be classified into LoS and NLoS components. Consequently, the radiated electric field of each patch antenna can be expressed as
\begin{equation} \label{equ:ComposedChannel}
\begin{aligned}
       {\mathbf{E}}_{\text{iso}}&=j\omega \mu \int \overline{\overline{\mathbf{G}}}_{\mathrm{S}}(\mathbf{r},\mathbf{r}') \cdot \mathbf{J}(\mathbf{r}') d\mathbf{r}'  = \bar {\mathbf{E}}_{\text{iso}} +  \tilde {\mathbf{E}}_{\text{iso}} ,
\end{aligned}
\end{equation}
where $ \bar {\mathbf{E}}_{\text{iso}}$ is the coherent propagation part and $\tilde {\mathbf{E}}_{\text{iso}}$ is the incoherent propagation part in the electric field.  

\subsubsection{Coherent Channel Modeling}
We first calculate the coherent part in the electric field (similar to LoS components), which is given by the real part of Green's functions $\bar{\bar{G}}_0(R)$ ($R$ is the distance between the central point of transmitter and receiver) \cite{10018012,8952896}.

The dyadic Green's function is given by \cite{arnoldus2001representation}
\begin{equation} \label{equ:dyadicGreen}
    \begin{aligned}
        \overline{\overline{\mathbf{G}}}_0(R)=&      \left(1-\frac{j}{kR} -\frac{1}{k^2 R^2} \right) \mathbf{I} \cdot \frac{\exp({-jkR})}{4\pi R}  \\ 
        &+ \left( \frac{3}{k^2R^2} 
        + \frac{3j}{kR} -1 \right) \vec{R} \vec{R}   \cdot        \frac{\exp({-jkR})}{4\pi R}  ,
    \end{aligned}
\end{equation}
where $\mathbf{I}$ is the identity matrix,  $\vec{R}$ is the direction of receive-transmit patch-antenna pair, and  
\begin{equation}
   \begin{aligned}
      \vec{R} \vec{R}   
       = \left[\begin{array}{ccc}
\cos x \cos x & \cos x \cos y  & \cos x \cos z  \\
\cos y \cos x & \cos y \cos y  & \cos y \cos z  \\
\cos z \cos x & \cos z \cos y  & \cos z \cos z  
\end{array}\right],
   \end{aligned} 
\end{equation}
where $\cos x=\frac{x_r-x_t}{R}, \cos y=\frac{y_r-y_t}{R}$, and $\cos z=\frac{z_r-z_t}{R}$, $(x_r,y_r,z_r)$ and $(x_t,y_t,z_t)$ are coordinates of receiver and transmitter, respectively.

Thus, the LoS channel is modeled as 
\begin{equation}
   \begin{aligned}
       \bar {\mathbf{E}}_{\text{iso}} (R)&=j\omega \mu \int \operatorname{Re} \left[\bar{\bar{\mathbf{G}}}_0(R)\right] \cdot \mathbf{J}(\mathbf{r}') \mathrm{d} \mathbf{r}'  \\
&       = \left[\begin{array}{ccc}
\bar{E}_{\text{iso}}^{xx}(R) & \bar{E}_{\text{iso}}^{xy} (R)& \bar{E}_{\text{iso}}^{xz}(R) \\
\bar{E}_{\text{iso}}^{yx}(R)& \bar{E}_{\text{iso}}^{yy}(R)& \bar{E}_{\text{iso}}^{yz}(R) \\
\bar{E}_{\text{iso}}^{zx}(R) & \bar{E}_{\text{iso}}^{zy}(R) & \bar{E}_{\text{iso}}^{zz} (R)
\end{array}\right],
   \end{aligned} 
\end{equation}
where each element can be computed from \eqref{equ:dyadicGreen}.

 
\subsubsection{Incoherent Channel Modeling}
Then, we calculate the incoherent (diffusive) propagation part  $\tilde E_{\text{iso}}$ in the electric field  (similar to NLoS components), which is dependent on the stochastic Green's function $\tilde{\mathbf{G}}_{\mathrm{S}}$ \cite{8952896}. 



The incoherent radiation field is 
\begin{equation}
    \begin{aligned}
       \tilde {\mathbf{E}}_{\text{iso}}=j \omega \mu  \int \tilde{\mathbf{G}}_{\mathrm{S}}\cdot \mathbf{J}(\mathbf{r}') \mathrm{d} \mathbf{r}' ,
    \end{aligned}
\end{equation}
with the stochastic Green's function given by 
\begin{equation}
    \begin{aligned}
        \tilde{\mathbf{G}}_{\mathrm{S}} = \sum_i \frac{\overline{\overline{\mathbf{D}}}(\mathbf{r},\mathbf{r}',k_i)}{\tilde{k}^2-k_i^2} ,
    \end{aligned}
\end{equation}
where $\tilde{k}=k(1-\frac{j}{2Q})$, and 
\begin{equation}
    \begin{aligned}
&\overline{\overline{\mathbf{D}}}\left(\mathbf{r}, \mathbf{r}^{\prime} ; k_i\right)=  D_{xx} \hat{\mathbf{x}} \hat{\mathbf{x}}+D_{xy} \hat{\mathbf{x}} \hat{\mathbf{y}}+D_{xz} \hat{\mathbf{x}} \hat{\mathbf{z}}+D_{yx}\hat{\mathbf{y}} \hat{\mathbf{x}}\\
&\quad +D_{yy} \hat{\mathbf{y}} \hat{\mathbf{y}}  +D_{yz} \hat{\mathbf{y}} \hat{\mathbf{z}}  +D_{zx}\hat{\mathbf{z}} \hat{\mathbf{x}}+D_{zy} \hat{\mathbf{z}} \hat{\mathbf{y}}+D_{zz} \hat{\mathbf{z}} \hat{\mathbf{z}} .
\end{aligned}
\end{equation}

Next, we will compute the covariance of each entry in $\overline{\overline{\mathbf{D}}}(\mathbf{r},\mathbf{r}',k_i)$.
\subsubsection{Copolarization components}
The expectation value of $D_{xx}$ and $D_{yy}$ is 
\begin{equation}
    \begin{aligned}
       &\quad \mathbb{E}[D_{xx}]= \mathbb{E}[D_{yy}]=\mathbb{E}\left[  {\Psi}_i^x(\mathbf{r})  {\Psi}_i^x(\mathbf{r}') \right]\\
       &=\frac{1}{8\pi^2 V} \int_0^{2\pi} \cos^2\psi_n \mathrm{d}\psi_n \cdot  \int_0^{2\pi} \sin^2 \phi_n \mathrm{d} \phi_n \\
       &\quad \cdot \int_0^{\pi} \sin \theta_n  \cos (k_i R \cos \theta_n) \mathrm{d} \theta_n \\
       &\quad+ \frac{1}{8\pi^2 V} \int_0^{2\pi} \sin^2\psi_n \cdot \int_0^{2\pi} \cos^2\phi_n \\
       &\quad   \cdot \int_0^{\pi} \sin\theta_n  \cos^2 \theta_n  \cos (k_i R \cos \theta_n) \mathrm{d} \theta_n \\
       &=\frac{1}{ 2V} \left[\frac{ \sin (k_i R)}{k_i R} + \frac{ \cos (k_i R)}{(k_i R)^2}-\frac{   \sin (k_i R) }{(k_i R)^3} \right],
    \end{aligned} 
\end{equation} 
which is obtained using the sum and product formulae of trigonometric identities.  

\color{black}

The second moment of $D_{xx}$ and $D_{yy}$ are  
\begin{equation}
    \begin{aligned}
       &\mathbb{E}[D_{xx}^2]=\mathbb{E}[D_{yy}^2] = \mathbb{E}[(\mathbf{\Psi}_i^x(\mathbf{r})  \mathbf{\Psi}_i^x(\mathbf{r}'))^2]  \\ 
       &=\frac{1}{8\pi^2 V^2}  \int_0^{2\pi} \cos^4\psi_n \mathrm{d}\psi_n \cdot   \int_0^{2\pi} \sin^4 \phi_n \mathrm{d} \phi_n \\
       &\quad \cdot  \int_0^{\pi} \sin \theta_n  \cos^2 (k_i R \cos \theta_n) \mathrm{d} \theta_n\\
       &\quad +\frac{1}{8\pi^2 V^2}   \int_0^{2\pi} \sin^4\psi_n \mathrm{d}\psi_n \cdot  \int_0^{2\pi} \cos^4 \phi_n \mathrm{d} \phi_n \\
       &\quad \cdot  \int_0^{\pi} \sin \theta_n \cos^2\theta_n \cos^2 (k_i R \cos \theta_n) \mathrm{d} \theta_n\\ 
       &=\frac{9}{128  V^2} \cdot \left[  \frac{4}{3}+ 2\frac{\sin (2k_i R)}{2k_i R} + \frac{2\cos (2k_i R)}{(2k_i R)^2} \right. \\
       &\quad \left.- \frac{2\sin (2k_i R)}{(2k_i R)^3}  \right].
    \end{aligned}
\end{equation}


Since the variance of variable $x$ is given by $\mathbb{V}[x]=\mathbb{E}[x^2]-\mathbb{E}[x]^2$, the variance of $D_{xx}$ and $D_{yy}$ is computed as

\begin{equation}
    \begin{aligned}
       & \mathbb{V}[D_{xx}]=\mathbb{V}[D_{yy}]\\
       &=\!\frac{9}{128  V^2} \! \left[ \! \frac{4}{3}\!+ \!2\frac{\sin (2k_i R)}{2k_i R} \!+\! \frac{2\cos (2k_i R)}{(2k_i R)^2} \! -\! \frac{2\sin (2k_i R)}{(2k_i R)^3}  \!\right]\\
       &\quad -\frac{1}{ 4V^2} \left[\frac{ \sin (k_i R)}{k_i R} + \frac{  k_i R \cos (k_i R) -  \sin (k_i R) }{(k_i R)^3} \right]^2.
    \end{aligned}
\end{equation}

Similarly, the expectation value of $D_{zz}$ is 
\begin{equation}
    \begin{aligned}
       &\quad \mathbb{E}[D_{zz}]=\mathbb{E}\left[\mathbf{\Psi}_i^z(\mathbf{r}) \mathbf{\Psi}_i^z(\mathbf{r}') \right] \\
       &=\frac{1}{4\pi V} \int_0^{2\pi} \! \! \sin^2\psi_n \mathrm{d}\psi_n  \! \int_0^{\pi} \! \! \sin^3 \theta  \cos (k_i R \cos \theta_n) \mathrm{d} \theta_n  \\
       &=\frac{1}{ V}  \frac{ \sin(k_i R)-  k_i R \cos (k_i R)}{(k_i R)^3} .
    \end{aligned} 
\end{equation}

The second moment of $D_{zz}$ is
\begin{equation} 
    \begin{aligned}
       &\mathbb{E}[D_{zz}^2]=\mathbb{E}[(\mathbf{\Psi}_i^z(\mathbf{r}) \mathbf{\Psi}_i^z(\mathbf{r}'))^2]  \\
       &=\frac{1}{4\pi V^2} \int_0^{2\pi}   \sin^4 \psi_n \mathrm{d} \psi_n  \cdot \int_0^{\pi} \sin^5 \theta_n \cos^2 \left( k_i R \cos \theta_n \right) \mathrm{d} \theta \\
       &=\frac{3}{16V^2} \left[ \int_{-1}^{1} \cos^2(k_i R u)\mathrm{d} u  - \int_{-1}^{1} 2 u^2 \cos^2(k_i R u)\mathrm{d} u \right. \\ 
       &\qquad \left.+ \int_{-1}^{1} u^4 \cos^2(k_i R u)\mathrm{d} u \right]\\
       &=\frac{3}{16V^2} \! \! \left[ \frac{8}{15} \!\! - \!\frac{8\sin(2k_i R)}{(2k_i R)^3} \! -\!  \frac{24\cos(2k_i R)}{(2k_i R)^4} \! + \! \frac{24\sin(2k_i R)}{(2k_i R)^5}  \right].
    \end{aligned}
\end{equation}

Therefore, the variance of  $D_{zz}$  is 
\begin{equation}
    \begin{aligned}
        &\mathbb{V} \left[D_{zz}\right]  = \frac{3}{16  V^2} \left[ \frac{8}{15} -\frac{8\sin(2k_i R)}{(2k_i R)^3} -  \frac{24\cos(2k_i R)}{(2k_i R)^4} \right. \\
        &\quad \left.+\frac{24\sin(2k_i R)}{(2k_i R)^5}  \right] \!  - \! \frac{1}{V^2} \frac{\left[ \sin \left(k_i R\right)-k_i R \cos \left(k_i R\right) \right]^2 }{\left(k_i R\right)^6}.
    \end{aligned} 
\end{equation}

\subsubsection{Cross-polarization components}
The expectation of cross-polarization components in the incoherent part is
\begin{equation}
    \begin{aligned}
       &\quad \mathbb{E}[D_{xy}]=\mathbb{E}[D_{xz}]=\mathbb{E}[D_{yx}]=\mathbb{E}[D_{yz}]\\
       &=\mathbb{E}[D_{zx}]=\mathbb{E}[D_{zy}] =0
    \end{aligned} 
\end{equation}

 The variances of $D_{xy}$  and  $D_{yx}$ are  
 \begin{equation}
    \begin{aligned}
       &\mathbb{E}[D_{xy}^2] = \mathbb{E}[D_{yx}^2]   = \mathbb{E}[(\mathbf{\Psi}_i^x(\mathbf{r})  \mathbf{\Psi}_i^y(\mathbf{r}'))^2]  \\  
       &= \frac{1}{128 V^2} \left[ \frac{64}{15} +8 \frac{\sin(2k_i R)}{2k_i R}  + 16\frac{\cos(2k_i R)}{(2k_i R)^2} \right.   \\
        &\quad \left.  - 40\frac{\sin(2k_i R)}{(2k_i R)^3} -  72\frac{ \cos(2k_i R)}{(2k_i R)^4}  + 72\frac{ \sin(2k_i R)}{(2k_i R)^5}   \right],
    \end{aligned}
\end{equation}
with the details given in the Appendix.

The variances of $D_{xz}, D_{yz}, D_{zx}$ and $D_{zy}$ are 

\begin{equation}
    \begin{aligned}
         & \mathbb{E}[D_{xz}^2] = \mathbb{E}[D_{yz}^2] =\mathbb{E}[D_{zx}^2] = \mathbb{E}[D_{zy}^2]\\
       &= \mathbb{E}[(\mathbf{\Psi}_i^x(\mathbf{r}) \mathbf{\Psi}_i^z(\mathbf{r}'))^2]  \\  
       &=  \frac{1}{8 \pi^2 V^2}  \left[ \int_0^{2\pi} \frac{1}{4}\sin^2 (2\psi_n) \mathrm{d} \psi_n \int_0^{2\pi} \sin^2\phi_n \mathrm{d}\phi_n  \right.\\
       &\quad \int_0^{\pi} \sin ^3\theta_n  \cos^2 (k_i R \cos \theta_n) \mathrm{d} \theta_n \\
       & \quad+ \int_0^{2\pi} \sin^4 ( \psi_n) \mathrm{d} \psi_n   \int_0^{2\pi} \cos^2  \phi_n  \mathrm{d} \phi_n \\
       &\quad  \left. \int_0^{\pi} \sin^3 \theta_n \cos^2\theta_n \cos^2 (k_i R \cos \theta_n) \mathrm{d} \theta_n   \right]\\
        &=\frac{1}{8 V^2} \left[ \frac{4}{15} - 2 \frac{ \cos (2k_i R)}{(2k_i R)^2} + 8 \frac{\sin(2k_i R)}{(2k_i R)^3} \right.\\
        &\qquad+    \left.\frac{18\cos(2k_i R)}{(2k_i R)^4}  - \frac{18\sin(2k_i R)}{(2k_i R)^5}   \right].
    \end{aligned}
\end{equation}

So far, we have obtained the probability density function (pdf) of each element in $\bar{\bar{D}} (\mathbf{r},\mathbf{r}',k_i)$, then we will elaborate the pdf of the denominator that depends on eigenvalue $k_i$, which is characterized by cavity quality factor $\tilde{Q}$ and normalized eigenvalue spacing $\bar{\Delta}$ given by \cite{10018012}
\begin{equation}
\begin{aligned}
    \tilde{k}^2-k_i^2 &\approx\left(\frac{k^2-k_i^2}{\bar{\Delta} }-j \frac{k^2}{ \tilde{Q} \bar{\Delta} }-  \frac{k^2}{4 \tilde{Q}^2 \bar{\Delta} }\right) \bar{\Delta}   \approx\left(\tilde{\lambda}_i-j \alpha\right) \bar{\Delta},
\end{aligned}
\end{equation}
with
\begin{equation} \label{equ:lambda}
\begin{aligned}
    &\tilde{\lambda}_i=\frac{k^2-k_i^2}{\bar{\Delta} }-\frac{k^2}{4Q^2 \bar\Delta}\overset{  \text{high} \;  \tilde{Q}}{\approx} \frac{k^2-k_i^2}{\bar{\Delta} },\\
    &\alpha=  \frac{k^3 V}{2\pi^2 \tilde{Q}}, \quad \bar{\Delta }=\frac{2\pi^2}{kV}.
\end{aligned}
\end{equation}

If the number of eigenvalues $M$ goes to infinity, i.e., the rich-scattering environment, the sum of the denominator can be approximately expressed as \cite{10018012}
\begin{equation}
    \begin{aligned}
\sum \frac{1}{\tilde{\lambda}_i^2+\alpha^2} & \approx  \int_{-\infty}^{\infty} \frac{\mathrm{d} \tilde{\lambda}_i}{\tilde{\lambda}_i^2+\alpha^2}=\frac{\pi}{\alpha} =\frac{2\pi^3 \tilde{Q}}{k^3 V} .
\end{aligned}
\end{equation}
\color{black}

Thus, 
\begin{equation}
    \sum_m \frac{\overline{\overline{\mathbf{D}}}\left(\mathbf{r}, \mathbf{r}^{\prime} ; k_m\right)}{\tilde{\lambda}_m-j \alpha} \frac{k V}{2 \pi^2}\approx  \frac{ \pi  \tilde{Q}}{k^2} \overline{\overline{\mathbf{D}}} \left(\mathbf{r}, \mathbf{r}^{\prime} ; k \right) ,
\end{equation}
where each term is given 
\begin{equation}
    \begin{aligned}
        & D_{xx}=D_{yy}\sim \mathcal{N} \left( \mathbb{E}[D_{xx}], \mathbb{V}[D_{xx}]\right), \\
        & D_{zz} \sim \mathcal{N} \left( \mathbb{E}[D_{zz}], \mathbb{V}[D_{zz}]\right), \\
        &D_{xy}=D_{yx}\sim \mathcal{N} \left( 0, \mathbb{V}[D_{xy}]\right), \\
        &D_{zx}=D_{zy}=D_{xz}=D_{xy} \sim \mathcal{N} \left( 0, \mathbb{V}[D_{zx}]\right).
    \end{aligned}
\end{equation}

 Therefore, the channel component satisfies 
\begin{equation}\label{equ:distriChannel}
        \mathbf{E}_{\mathrm{iso}}=\bar{\mathbf{E}}_{\text{iso}} + \tilde{\mathbf{E}}_{\text{iso}}\sim \mathcal{N} \left( \mathbb{E}( \mathbf{E}_{\mathrm{iso}}), \mathbb{V}(\mathbf{E}_{\mathrm{iso}}) \right),
\end{equation}
with the mean $\mathbb{E}(\mathbf{E}_{\text{iso}})=\bar{\mathbf{E}}_{\text{iso}} + \tilde{\mathbf{E}}_{\text{iso}} $ and variance $\mathbb{V}(\mathbf{E}_{\text{iso}})=\mathbb{V}(\tilde{\mathbf{E}}_{\text{iso}})$. The means of LoS channel $\bar{\mathbf{E}}_{\text{iso}} $, NLoS channel $\tilde{\mathbf{E}}_{\text{iso}}$, and the variance of channel $\mathbb{V}(\mathbf{E}_{\text{iso}})$ in $(pq)$-th ($p,q\in \{x,y,z \}$) polarization state are given by 
\begin{equation}
    \begin{aligned}
    &\mathbb{E}(\bar{E}_{\text{iso}}^{pq})\!=j \omega \mu  I  \bar{E}_{\text{iso}}^{pq}(R),  \mathbb{E}(\tilde{E}_{\text{iso}}^{pq})\!= j \omega \mu  I \frac{ \pi  \tilde{Q}}{k^2} \mathbb{E} \left[ D_{pq} \right] ,\\
    & \mathbb{V}(E_{\text{iso}}^{pq})\!=  \omega^2 \mu^2 I^2 \frac{ \pi^2  \tilde{Q}^2}{k^4} \mathbb{V}\left[ D_{pq} \right]  .
    \end{aligned}
\end{equation}

Obviously, the cavity volume $V$ and quality factor $\tilde{Q}$ jointly have impacts on the LoS and NLoS channel contributions. To further simplify the channel expression, we employ the $K$-factor to control the ratio of the LoS component over the NLoS component \cite{6557073} for comparison with the conventional Rayleigh channel model. The simulated $K$-factor is related to three parameters: the cavity quality factor $\tilde{Q}$, cavity volume $V$, and distance $R$ between the transmitter and receiver, which is similar to the conclusion in \cite{4012433}. 

In order to further simplify the expression, we incorporate the joint impact of $\tilde{Q}$ and $V$ in the revised $K$-factor.  Specifically, the LoS component and NLoS component for $K$-factor are 
\begin{equation}
    \begin{aligned}
    & K_{\text{LoS}} = \frac{K}{c+K },\quad K_{\text{NLoS}}=\frac{ c }{c+K},
    \end{aligned}
\end{equation}
with the constant $c$ given by
\begin{equation}
    c= \frac{k^4}{\pi^2 \tilde{Q}^2} \cdot \frac{\mathbb{E}\left[\bar{E}_{\mathrm{iso}}^2\right]}{\mathbb{E}^2 \left[ D_{pq} \right] + \mathbb{V}\left[ D_{pq} \right] }.
\end{equation}

Then, the channel \eqref{equ:ComposedChannel} using $K$-factor representation can be rewritten as 
\begin{equation} \label{equ:KComposedChannel}
\begin{aligned}
       {\mathbf{E}}_{\text{iso}}& =\sqrt{\frac{K}{c+K }} \bar {\mathbf{E}}_{\text{iso}} +  \sqrt{\frac{ c }{c+K}} \tilde {\mathbf{E}}_{\text{iso}}.
\end{aligned}
\end{equation}  
 
\subsection{Mutual Coupling in HMIMO Systems} 
The mutual coupling matrix is dependent on the geometry of the antenna array and antenna element impedance. Typically, the coupling matrix $\mathbf{C}\in \mathbb{C}^{N\times N}$ ($N$ is the number of antennas) can be written using fundamental electromagnetics and circuit theory as  \cite{1167256,1291781,9512430}
\begin{equation}\label{equ:MC_all}
    \mathbf{C}=\mathbf{Z} (\mathbf{Z} +\mathbf{Z}_s \mathbf{I})^{-1},
\end{equation}
where $\mathbf{Z} \in\mathbb{C}^{N \times N}$ is the mutual impedance matrix obtained from EM simulation software CST. The diagonal elements represent the impedance of each antenna, and the off-diagonal elements represent the mutual coupling between the two antennas. If the antenna elements are uncoupled, the impedance matrix $\mathbf{Z}$ becomes diagonal. $\mathbf{Z}_s$ is the impedance of each element in the transmitter/receiver, and is chosen as the diagonal matrix of  $\mathbf{Z}$.

Based on \eqref{equ:MC_all}, the mutual coupling effects at the transmitter and receiver are modeled as  $\mathbf{C}_s\in \mathbb{C}^{N_s\times N_s}$ and $\mathbf{C}_r\in \mathbb{C}^{N_r\times N_r}$, respectively. Then, the mutual-coupling-induced channel is given by 
\begin{equation} \label{equ:scatterChannel}
    \begin{aligned}
        \mathbf{E} & =\mathbf{C}_r  \mathbf{E}_{\mathrm{iso}} \mathbf{C}_s\\
        &=\mathbf{C}_r \left[\begin{array}{ccc  }
				E_{\text{iso}} (R_{1,1})  &  \cdots &  E_{\text{iso}} (R_{1,N_s})   \\ 
				E_{\text{iso}} (R_{2,1})  &   \cdots  &  E_{\text{iso}} (R_{2,N_s})  \\
                                 & \ddots         &     \\
				 E_{\text{iso}} (R_{N_r,1})  &  \cdots  &  E_{\text{iso}} (R_{N_r,N_s})
			\end{array}\right] \mathbf{C}_s,
    \end{aligned}
\end{equation}
where each element of isolated channel matrix $\mathbf{E}_{\mathrm{iso}}\in \mathbb{C}^{N_r \times N_s}$ follows the distribution given by  \eqref{equ:distriChannel} and \eqref{equ:KComposedChannel}, and $R_{m,n}, m=1,\ldots,N_r, n=1,\ldots, N_s$ is the distance between the $m$-th receiver antenna and $n$-th transmit antenna.

\textit{Remarks: To ensure tolerable errors in the above channel modeling, the antenna elements must be approximated to minimum scatterers, i.e., the spacing between two adjacent antennas at the edge should not be less than about $\lambda/4$ \cite{kildal2015foundations,299601}. }

\color{black}
\subsection{Implications of Channels for HMIMO Systems}
Since HMIMO is composed of a large number of antenna elements, we consider an approximation method to compute the radiation intensity, total radiated power, and directivity gain. Specifically, if we assume that the mutual coupling only exists at the transmitter, and denote the distance information between the transmitter and receiver is collected within $\boldsymbol{R}=[R_{1,1},\ldots, R_{m,n},\ldots, R_{N_r, N_s}], m=1,\ldots,N_r,n=1,\ldots,N_s$. The current distribution at the transmitter is assumed to be the identity matrix, i.e., we only investigate the impact of channels here, then the radiation intensity is 
\begin{equation} \label{equ:RadiIntensity}
\begin{aligned}
    &U(\boldsymbol{R})=\sum_{m}\sum_n \mathbb{E} \left(  \mathbf{E}_{m,n}(\boldsymbol{R})\cdot \mathbf{E}_{m,n}^{*}(\boldsymbol{R}) \right) \\ 
   &=\sum_{m}\sum_n \left[ \mathbf{C}_s^{\dagger} \mathbb{E} \left[ \operatorname{diag} \left( \sum_{n=1}^{N_s}   E_{\text{iso}}^2(R_{1,n}), \ldots, \right. \right.\right. \\
   &  \sum_{n=1}^{N_s}   E_{\text{iso}}^2 \!(R_{m,n}),   
   \left. \left. \left.\ldots, \!\sum_{n=1}^{N_s}    E_{\text{iso}}^2\! (R_{N_r,n}) \right) \right]\! \mathbf{C}_s \right]_{mn}\!\!,  
\end{aligned}
\end{equation}
where 
\begin{equation}
    \begin{aligned}
        &\mathbb{E}\left[ \sum_{n=1}^{N_s}  E_{\text{iso}}^2(R_{m,n}) \right]= \sum_{n=1}^{N_s}\mathbb{E}  \left[  E_{\text{iso}}^2(R_{m,n}) \right]\\
        &=\sum_{n=1}^{N_s}  \mathbb{E}^2 \left[ E_{\mathrm{iso}} (R_{m,n})\right]  + \mathbb{V} [E_{\mathrm{iso}} (R_{m,n})].
    \end{aligned}
\end{equation}

The average radiation power $P_{\text{av}}$ with a given direction $\Omega$ is given by 
\begin{equation}
    \begin{aligned}
       & P_{\text{av}}=  \mathbb{E} ^2 \left[E_{\mathrm{iso}}(R_{o,o})  \right] \|\mathbf{C}_s \|^2  \\
        &=\sum_{m}\sum_n \left[ \mathbf{C}_s^{\dagger}    \operatorname{diag} \left( \sum_{n=1}^{N_s} \mathbb{E}^2 \left[ E_{\text{iso}}(R_{1,n})\right], \ldots, \right. \right.  \\
   &   \sum_{n=1}^{N_s}   \mathbb{E}^2 \left[ E_{\text{iso}}(R_{m,n})\right]  ,   
   \left. \left. \! \!\ldots, \! \sum_{n=1}^{N_s} \mathbb{E}^2  \!\left[ E_{\text{iso}}(R_{N_r,n})\right] \! \right)  \!\mathbf{C}_s \! \right]_{mn}\!\!.
    \end{aligned}
\end{equation}

Antenna gain, which describes the ability to convert input power into radiation in a specified direction $\hat{\mathbf{r}}$ for HMIMO systems, is proportional to the quotient between the radiation intensity $U(\boldsymbol{R})$ and the dissipated power $P_r+P_{\Omega}$, i.e. \cite{8718501},
\begin{equation}
    G(\boldsymbol{R})=4\pi \frac{U(\boldsymbol{R})}{P_r+P_{\Omega}},
\end{equation}
where $P_r$ is the radiated power, and $P_{\Omega}$ is the power dissipated in ohmic and dielectric losses. If we assume there is no loss, i.e., $P_{\Omega}=0$, where the gain array gain equals to the directivity $D(\boldsymbol{R})$, then we have 
\begin{equation} \label{equ:direct}
\begin{aligned}
        G(\boldsymbol{R})&=D(\boldsymbol{R})=4\pi \frac{U(\boldsymbol{R})}{P_{\text{av}}}.
\end{aligned}
\end{equation}

\section{ Theoretical Capacity Evaluation}
In this section, we will evaluate theoretical capacity bounds for SISO systems and MISO systems in different communication regions. In addition, the capacity region for two-user systems in various communication zones is also presented. 
\subsection{SISO Systems}
Assume the single-antenna transmitter is located at $\mathbf{r}_t=(x_t,y_t,z_t)$, the single-antenna receiver is located at $\mathbf{r}_r=(x_r,y_r,z_r)$, and the distance between the transmitter and receiver is $R=|\mathbf{r}_s-\mathbf{r}_r|$.   Based on the proposed channel model, the channel of SISO systems is composed of coherent part $\bar{\mathbf{H}}_{\text{iso}}^{\text{SISO}}(R)$ and incoherent part $\tilde{\mathbf{H}}_{\text{iso}}^{\text{SISO}}(R)$, i.e., 
\begin{equation}  
\mathbf{H}^{\text{SISO}}\left(R\right)=\bar{\mathbf{H}}_{\text{iso}}^{\text{SISO}}(R)+\tilde{\mathbf{H}}_{\text{iso}}^{\text{SISO}}(R).
\end{equation}

For notation simplicity, the constant $j \omega \mu  I$ is omitted in the following. The coherent channel is   
\begin{equation}  
    \begin{aligned}
       & \bar{\mathbf{H}}_{\text{iso}}^{\text{SISO}}(R) \propto 
       \mathrm{Re}\!\left[\bar{\bar{\mathbf{G}}}_0(R) \right]  \\ & =\left(\frac{\cos(kR)}{4\pi R} \! - \!\frac{\sin(kR)}{4\pi k R^2} \!- \!\frac{\cos{(kR)}}{4\pi k^2 R^3} \right) \mathbf{I} \\
        &\quad  +  \left( \frac{3\cos(kR)}{4\pi k^2R^3}   + \frac{3 \sin(kR)}{4\pi kR^2} -  \frac{\cos(kR)}{4\pi R}  \right) \vec{R} \vec{R} .\\
    \end{aligned}
\end{equation}

The incoherent channel is given by 
\begin{equation}
    \begin{aligned}
     \mathbb{E}\! \left[\tilde{\mathbf{H}}_{\text{iso}}^{\text{SISO}}(R)\right]\!\! \propto \!  \! \frac{ \pi  \tilde{Q}}{k^2} \mathbb{E} \left[ \mathbf{D} \right] ,  \mathbb{V}\! \left[\tilde{\mathbf{H}}_{\text{iso}}^{\text{SISO}}(R) \right]\!\! \propto \!  \! \frac{ \pi^2  \tilde{Q}^2}{k^4} \mathbb{V}\left[ \mathbf{D} \right]  .
    \end{aligned}
\end{equation}

Take the single polarized channel (e.g., XX/YY-th polarization) as an example, the expectation of the SISO channel is
\begin{equation}
    \begin{aligned}
       &\quad \mathbb{E}[H^{\text{SISO}}_{xx}]= \mathbb{E}[H^{\text{SISO}}_{yy}] \\
       & \propto \frac{\cos(kR)}{4\pi R} \! - \!\frac{\sin(kR)}{4\pi k R^2} \!- \!\frac{\cos{(kR)}}{4\pi k^2 R^3}  \\
        &\quad +  \left[ \frac{3\cos(kR)}{4\pi k^2R^3}   + \frac{3 \sin(kR)}{4\pi kR^2} -  \frac{\cos(kR)}{4\pi R}  \right] \cos x \cos x  \\
        &\quad +\frac{\pi  \tilde{Q} }{ 2Vk^2} \left[\frac{ \sin (k  R)}{k  R} + \frac{ \cos (k  R)}{(k  R)^2}-\frac{   \sin (k  R) }{(k  R)^3} \right], 
\end{aligned} 
\end{equation}
where the first two terms are coherent channel components, and the last term is an incoherent channel component. 

The variance of the SISO channel  in the XX/YY-th polarization state is
\begin{equation} \label{equ:SISO_var}
    \begin{aligned}        
       & \mathbb{V}[H^{\text{SISO}}_{xx}]=\mathbb{V}[H^{\text{SISO}}_{yy}]\\
       & \propto\frac{9\pi^2  \tilde{Q}^2}{128  V^2 k^4}  \left[  \frac{4}{3}+ 2\frac{\sin (2k  R)}{2k  R} + \frac{2\cos (2k  R)}{(2k  R)^2}  - \frac{2\sin (2k  R)}{(2k  R)^3}  \right]\\
       &\quad -\frac{\pi^2  \tilde{Q}^2}{ 4V^2 k^4} \left[\frac{ \sin (k  R)}{k  R} + \frac{  k  R \cos (k  R) -  \sin (k  R) }{(k  R)^3} \right]^2.
    \end{aligned} 
\end{equation}

Therefore, the channel capacity of single-polarized SISO systems (the subscript is omitted) is given by \cite{1237143, 850678,7024192}
\begin{equation}
    \begin{aligned}
        \mathcal{C}^{\text{SISO}}&=\mathbb{E}\left[ \log_2\left( 1+ |H^{\text{SISO}} |^2/\sigma^2 \right) \right]\\
        &\geq \log_2\left( 1+ \mathbb{E}\left[|H^{\text{SISO}} |^2\right]/\sigma^2 \right) \\
        &= \log_2\left( 1+ \left(\mathbb{E}[H^{\text{SISO}} ]^2 +  \mathbb{V}[H^{\text{SISO}} ]\right) /\sigma^2 \right) ,
    \end{aligned}
\end{equation}
where $\sigma^2$ is the variance of additive noise and the Jensen's inequality is adopted. 

Typically, in the far-field region where the terms with $R^2$ and $R^3$ in denominators can be omitted, the single-polarized SISO channel is approximated by 
\begin{equation} \label{equ:SISO_var_ff}
    \begin{aligned}
       & \mathbb{E}[H^{\text{SISO}} ] \!  \propto \!\frac{\cos(kR)}{4\pi R}     +\frac{\pi  \tilde{Q}  \sin (k  R)}{ 2Vk^3 R}   , \\     
       & \mathbb{V}[H^{\text{SISO}} ] \! \propto \!\frac{9\pi^2  \tilde{Q}^2}{128  V^2 k^4}  \left[  \frac{4}{3}\!+\! 2\frac{\sin (2k  R)}{2k  R}    \right] \!  - \! \frac{\pi^2  \tilde{Q}^2}{ 4V^2 k^4} \left[\frac{ \sin (k  R)}{k  R}  \right]^2 \!.
    \end{aligned} 
\end{equation}

The capacity bound is given by 
\begin{equation}
    \begin{aligned}
        &\mathcal{C}^{\text{SISO}}_{\text{far-field}} \geq   \log_2\left\{ 1+ \left(\mathbb{E}[H^{\text{SISO}} ]^2 +  \mathbb{V}[H^{\text{SISO}} ]\right) /\sigma^2 \right\} \\
        &= \log_2\left\{ 1+\! \!\left(\frac{\cos^2(kR)}{16\pi^2 R^2} \! + \! \frac{\tilde{Q}}{4kV}\! \cdot\! \frac{\cos(kR)}{  k R}\! \cdot \! \frac{    \sin (k  R)}{  k  R}\right.\right. \\
        & \quad \left. \left. +   \frac{3\pi^2  \tilde{Q}^2}{32  V^2 k^4}+ \frac{9\pi^2  \tilde{Q}^2}{64  V^2 k^4}  \cdot \frac{\sin (2k  R)}{2k  R} \right) /\sigma^2 \right\}  .
    \end{aligned}
\end{equation} 

Then, the asymptotic capacity analysis of SISO systems is extended to MISO systems. 

\subsection{MISO Systems}
Assume the transmitter is equipped with $N_s$ antennas that are linearly located along the $\hat{\mathbf{x}}$ direction with spacing $\Delta_s$, and the location of the $n$-th antenna is denoted as $\mathbf{r}_{t,n}=((n-1)\Delta_s,0,0), n=1,\ldots,N_s$. The receiver with the single antenna is located at $\mathbf{r}_r=(0,0,R)$.  Therefore, the distance between the receiver and the $n$-th transmitter is $R_n=\sqrt{R^2+(n-1)^2\Delta_s^2}, i=1,\ldots, N_s$.

Since we only consider the single-polarization state, the $n$-th component in the MISO channel $\mathbf{H}^{\text{MISO}}=[H^{\text{MISO}}_1,\ldots, H^{\text{MISO}}_{N_s} ] \in \mathbb{C}^{1\times N_s}$ is 
\begin{equation}
    \begin{aligned}
        &\mathbb{E}[H^{\text{MISO}}_{n}]  \propto \frac{\cos(kR_n)}{4\pi R_n} \! - \!\frac{\sin(kR_n)}{4\pi k R_n^2} \!- \!\frac{\cos{(kR_n)}}{4\pi k^2 R_n^3}  \\
        &\quad \! + \! \left[ \frac{3\cos(kR_n)}{4\pi k^2R_n^3}   \!+ \!\frac{3 \sin(kR_n)}{4\pi kR_n^2}\! - \! \frac{\cos(kR_n)}{4\pi R_n}  \right]\! \cos x_n \!\cos x_n  \\
        &\quad +\frac{\pi  \tilde{Q} }{ 2Vk^2} \left[\frac{ \sin (k  R_n)}{k  R_n} + \frac{ \cos (k  R_n)}{(k  R_n)^2}-\frac{   \sin (k  R_n) }{(k  R_n)^3} \right], 
\end{aligned} 
\end{equation}
where $\cos x_n=\frac{(n-1)\Delta}{R_n}\!=\!\frac{(n-1)\Delta_s}{\sqrt{R^2+(n-1)^2\Delta_s^2}}\!=\!\frac{1}{\sqrt{\left( \frac{R}{(n-1)\Delta_s}\right)^2+1}}$.

 The variance of the $n$-th component in the MISO channel is given in \eqref{equ:SISO_var} with $R=R_n$.

Similarly, the channel capacity of single-polarized MISO systems is given by
\begin{equation} \label{equ:MISO_cap}
    \begin{aligned}
        \mathcal{C}^{\text{MISO}}&=\mathbb{E}\left[ \log_2\left( 1+  \mathbf{H}^{\text{MISO}} {\mathbf{H}^{\text{MISO}}}^{\dagger}  /\sigma^2 \right) \right]\\
        &\geq \log_2\left( 1+ \mathbb{E}\left[ \mathbf{H}^{\text{MISO}} {\mathbf{H}^{\text{MISO}}}^{\dagger} \right]/\sigma^2 \right) ,
    \end{aligned}
\end{equation}
with 
\begin{equation}
    \begin{aligned}
        \mathbb{E} \left[ \mathbf{H}^{\text{SISO}} {\mathbf{H}^{\text{SISO}}}^{\dagger}\right] 
       =\sum_{n=1}^{N_s}  \mathbb{E}[H^{\text{MISO}}_n ]^2 +  \mathbb{V}[H^{\text{MISO}}_n ] .
    \end{aligned}
\end{equation}

Different from SISO systems, the channel capacity of MISO systems is jointly impacted by distance $R$ and aperture size $\mathcal{A}_s=(N_s-1)\Delta_s$. Consequently, we investigate the following special cases:
\begin{itemize}
    \item \textbf{Case 1 (Lower bound):} The terms $\frac{1}{R^2}\rightarrow 0$ and $\frac{1}{R^3}\rightarrow 0$ in the middle/far-field region, thus yielding the lower bound of the capacity of the MISO systems. 
    
Specifically, the expectation of the $n$-th component in the MISO channel is approximated by 
\begin{equation}
    \begin{aligned}
        &\mathbb{E}[H^{\text{MISO}}_{n}]  \propto \frac{\cos(kR_n)}{4\pi R_n}   +\frac{\pi  \tilde{Q} }{ 2Vk^2}  \frac{ \sin (k  R_n)}{k  R_n}  \\
        &\qquad  \qquad \quad -  \frac{\cos(kR_n)}{4\pi R_n} \frac{(n-1)^2\Delta_s^2}{(n-1)^2\Delta_s^2+R^2} \\
        &\qquad  \qquad =\frac{\cos(kR_n)}{4\pi R_n} \frac{R^2}{R_n^2} +\frac{\pi  \tilde{Q} }{ 2Vk^2}  \frac{ \sin (k  R_n)}{k  R_n}  , 
\end{aligned} 
\end{equation}
and the variance approximation $\mathbb{V}[H^{\text{MISO}} ]$ is given in \eqref{equ:SISO_var_ff} with $R=R_n$. 

Therefore, the squares of $\mathbb{E}[H^{\text{MISO}}_{n}]$ is given by 
\begin{equation} \label{equ:expNF}
    \begin{aligned}
        &\mathbb{E}^2[H^{\text{MISO}}_{n}] \! \propto \!\left[\frac{\cos(kR_n)}{4\pi R_n} \frac{R^2}{R_n^2} \right]^2 \!  + \! \left[ \frac{\pi  \tilde{Q} }{ 2Vk^2}  \frac{ \sin (k  R_n)}{k  R_n}\right]^2 \\
        & \quad +  \frac{  \tilde{Q} }{ 4Vk }  \frac{\cos(kR_n)}{k R_n} \frac{ \sin (k  R_n)}{k  R_n}  \frac{R^2}{R_n^2}, 
\end{aligned} 
\end{equation}
which can be adopted in \eqref{equ:MISO_cap} to calculate the lower bound of channel capacity $\mathcal{C}^{\text{MISO}}$ given by 
\begin{equation}
    \begin{aligned} \label{equ:LowerBound}
        &\mathcal{C}^{\text{MISO}}_{\text{lower-bound}} \!\geq \!   \log_2\left\{  \!1+\! \!\left( \mathbb{E}^2[H^{\text{MISO}}_{n}]   +   \mathbb{V}[H^{\text{MISO}} ] \right) /\sigma^2 \right\}  .
    \end{aligned}
\end{equation} 

\item \textbf{Case 2 (Upper bound):} Since the smaller distance between the transmitter and receiver antennas results in a higher capacity, the upper bound of the MISO channels can be obtained for $R_n\approx R, n=1,\ldots, N_s$, which is achievable for a negligible antenna aperture compared with the transmitter-receiver distance, i.e., $(N_s-1)\Delta_s\ll R$ and $R_n\approx R$. 

Therefore, the capacity of the MISO channel is upper-bounded by  
\begin{equation}
    \begin{aligned} \label{equ:UpperBound}
        &\mathcal{C}^{\text{MISO}}_{\text{upper-bound}} \!\leq\! \! \log_2\left\{ \!1+ \!\!\left(\!\sum_{n=1}^{N_s} \! \mathbb{E}^2[H^{\text{MISO}}_n ] \!+ \! \mathbb{V}[H^{\text{MISO}}_n ] \!\right)\!/\!\sigma^2 \!\right\} \\
        &=\log_2\left\{ 1 \! + \!N_s \left(\frac{\cos^2(kR)}{16\pi^2 R^2} \! + \! \frac{\tilde{Q}}{4kV} \!\cdot \!\frac{\cos(kR)}{  k R} \!\cdot \!\frac{    \sin (k  R)}{  k  R}\right.\right. \\
        & \quad \left. \left. \!+ \!  \frac{3\pi^2  \tilde{Q}^2}{32  V^2 k^4}+ \frac{9\pi^2  \tilde{Q}^2}{64  V^2 k^4}  \cdot \frac{\sin (2k  R)}{2k  R} \right) /\sigma^2 \right\}  ,
    \end{aligned}
\end{equation} 
where the $N_s$ antennas in the transmitter equally contribute to the capacity improvement of MISO systems in far-field regions. 

\end{itemize}

To sum up, the lower-bounded capacity $\mathcal{C}_{\text{lower-bound}}^{\text{MISO}}$ in \textbf{Case 1} approximates the capacity of MISO systems well in the near/middle field and underestimates greatly in the far field, while the upper-bounded capacity $\mathcal{C}^{\text{MISO}}_{\text{upper-bound}} $ in \textbf{Case 2} is similar to that in the conventional i.i.d.  Rayleigh channel assumption, thus approaching the capacity of practical MISO systems in the far-field region.  

\subsection{Capacity Region for Two-User Case}
Considering a two-user case with channels $\mathbf{H}^{(1)}$ and $\mathbf{H}^{(2)}$, where the symbols  $\mathbf{X}^{(1)}$ and $\mathbf{X}^{(2)}$ sent by the first user and second user are modulated by the current distribution, respectively. The input-output relationship is given by
\begin{equation}
    \mathbf{Y}=\mathbf{X}^{(1)}  \mathbf{H}^{(1)} + \mathbf{X}^{(2)}  \mathbf{H}^{(2)} + \mathbf{N},
\end{equation}
where  $\mathbf{Y}$ is the output signal at the observed field, and $\mathbf{N}$ is the additive noise with zero means and unit variances. 

Thus, the capacity region \cite{1237406} is 
\begin{equation}\label{equ:capacityRegion}
    \begin{aligned}
        \mathcal{C}_{G}\!\!=    \! \bigcup  & \left\{ (\mathcal{R}^{(1)},\mathcal{R}^{(2)}):  \right.  \\
        & \quad 0\leq \mathcal{R}^{(1)}\leq  \mathcal{I}\left( \mathbf{X}^{(1)};\mathbf{Y}   \mid \mathbf{X}^{(2)} \right) \\ 
        & \quad  0\leq \mathcal{R}^{(2)}\leq   \mathcal{I}\left( \mathbf{X}^{(2)};\mathbf{Y}   \mid \mathbf{X}^{(1)} \right) \\
        & \quad  0\leq \mathcal{R}^{(1)}+\mathcal{R}^{(2)}\leq  \mathcal{I}\left( \mathbf{X}^{(1)},\mathbf{X}^{(2)};\mathbf{Y}   \right)  \left. \right \},
    \end{aligned}
\end{equation} 
with
\begin{equation}
\begin{aligned}
     \mathcal{I}(\mathbf{X},\mathbf{Y} )\! = \! \mathcal{H}(\mathbf{Y} )\!-\!\mathcal{H}(\mathbf{Y} \mid \mathbf{X}) \!= \! \log_2 \left | \mathbf{I} \!+\! \frac{ \mathbf{H}  \mathbf{W}_{xx} \mathbf{H}  ^{\dagger} } { \sigma ^2}\right  |,
\end{aligned}
\end{equation}
where $\mathcal{H}(\cdot)$ denotes the entropy, $\mathcal{I}(\mathbf{X},\mathbf{Y} )$ is the mutual information between the input signal $\mathbf{X}$ and output signal $\mathbf{Y} $, and the covariance matrix of current distribution $\mathbf{W}_{xx}=\mathbb{E}[\mathbf{X}\mathbf{X}^{\dagger}]=\mathbf{I}$ is assumed for simplicity. 

Typically, the capacity region is convex, and the boundary of the capacity region is obtained by solving optimization problems that are parameterized by the slope of the supporting hyperplanes \cite{6503473}.  There are two corner points (A and B) in the capacity region, and each corresponds to the sum rate with perfect successive interference cancellation. Specifically, the capacity of user 1, $\mathcal{R}^{(1)}$, reduces from $\log_2(\mathbf{I}+\frac{1}{\sigma ^2} {\mathbf{H}^{(1)}}^{\dagger} \mathbf{H}^{(1)})$ to  $\log_2(\mathbf{I}+\frac{1}{\sigma ^2} \mathbf{H}^{\dagger} \mathbf{H})-\log_2(\mathbf{I}+\frac{1}{\sigma ^2} {\mathbf{H}^{(1)}}^{\dagger} \mathbf{H}^{(1)})$, and the capacity of user 2 $\mathcal{R}^{(2)}$ increases from   $\log_2(\mathbf{I}+\frac{1}{\sigma ^2} \mathbf{H}^{\dagger} \mathbf{H})-\log_2(\mathbf{I}+\frac{1}{\sigma ^2} {\mathbf{H}^{(2)}}^{\dagger} \mathbf{H}^{(2)})$ to $\log_2(\mathbf{I}+\frac{1}{\sigma ^2} {\mathbf{H}^{(2)}}^{\dagger} \mathbf{H}^{(2)})$.

Here, we investigate the theoretical capacity region in different communication regions. 
\begin{itemize}
    \item \textbf{Case 3 (two near/middle-field users):} Since the two users are located in the middle/near-field region, the achievable capacity can be approximated by \eqref{equ:LowerBound}. Thus, the corner point for the $i$-th user is approximated by:
     \begin{equation}
         \begin{aligned}
             \mathcal{R}^{(i)} \approx    \log_2\left\{ 1+\! \!\left( \mathbb{E}^2[H^{(i)}_{n}]   +   \mathbb{V}[H^{(i)} ] \right) /\sigma^2 \right\} ,
         \end{aligned}
     \end{equation}
     where $H^{(i)}_{n}$ is the $n$-th entry in channel vector $\mathbf{H}^{(i)}$, the squares of expectation are given in \eqref{equ:expNF} and the variances are given in \eqref{equ:MISO_cap}.

     \item \textbf{Case 4 (two far-field users):} As the two users are located in the far-field region, the achievable capacity of each can be approximated by \eqref{equ:UpperBound}. Thus, the corner point for the $i$-th user is approximated by:
     \begin{equation}
         \begin{aligned}
             & \mathcal{R}^{(i)}\! \approx \!  \log_2\left\{ 1\!+ \!N_s \left(\frac{\cos^2(kR^{(i)})}{16\pi^2  {R^{(i)}}^2}  \!+ \! \frac{\tilde{Q}}{4kV}\! \cdot \!\frac{\cos(k{R^{(i)}})}{  k {R^{(i)}}} \right.\right. \\
             &   \left. \left. \cdot \frac{    \sin (k  {R^{(i)}})}{  k  {R^{(i)}}} \! +  \! \frac{3\pi^2  \tilde{Q}^2}{32  V^2 k^4} \!+\! \frac{9\pi^2  \tilde{Q}^2}{64  V^2 k^4} \! \cdot\! \frac{\sin (2k  {R^{(i)}})}{2k  {R^{(i)}}}\! \right) \!/\!\sigma^2 \!\right\}  , 
         \end{aligned}
     \end{equation}
     where ${R^{(i)}}$ is the distance from the transmitter to the central point in the $i$-th user.  

     \item \textbf{Case 5 (one far-field user and one near/middle-field user):} As the two users are located in distinct communication regions, the achievable capacity of the far-field user can be approximated by \eqref{equ:UpperBound}, while that of the near-field user can be approximated by \eqref{equ:LowerBound}.  
\end{itemize} 
In all cases, the middle points along the line between two corner points are given by the combination of $\mathcal{R}^{(1)} $ and $\mathcal{R}^{(2)} $, i.e., $\epsilon \mathcal{R}^{(1)} + (1-\epsilon) \mathcal{R}^{(2)} $, $\epsilon\in[0,1]$.

\section{Simulation Results}
\begin{figure*}[htbp] 
		\centering  
		\subfigure[$d_{\mathrm{RT}}=0.4\lambda$]{
    \begin{minipage}[t]{0.45\linewidth}
        \centering
        \includegraphics[width=1\linewidth]{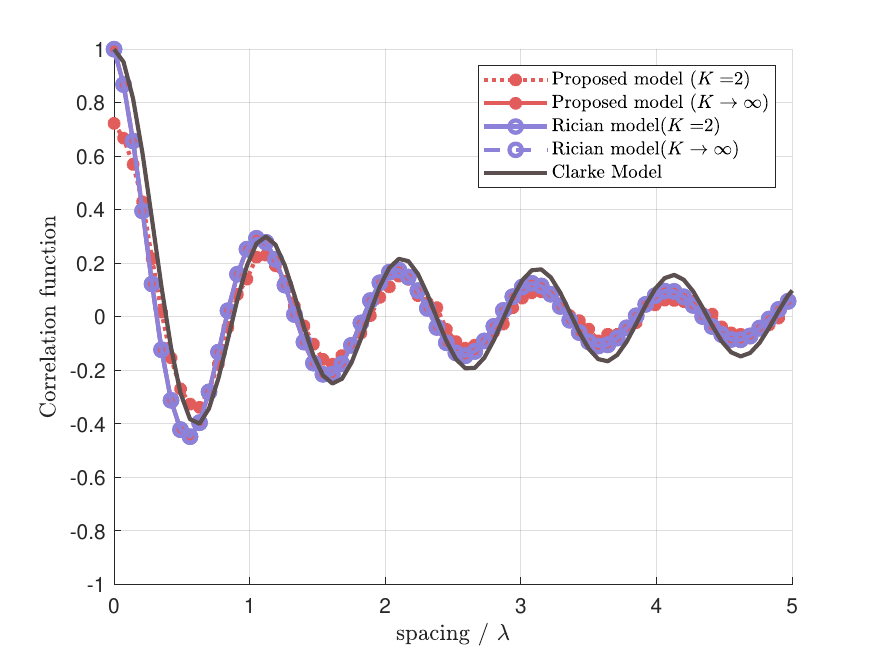}  
        \vspace{0.02cm}  
    \end{minipage}%
} 
\subfigure[$d_{\mathrm{RT}}=10\lambda$]{
    \begin{minipage}[t]{0.45\linewidth}
        \centering
        \includegraphics[width=1\linewidth]{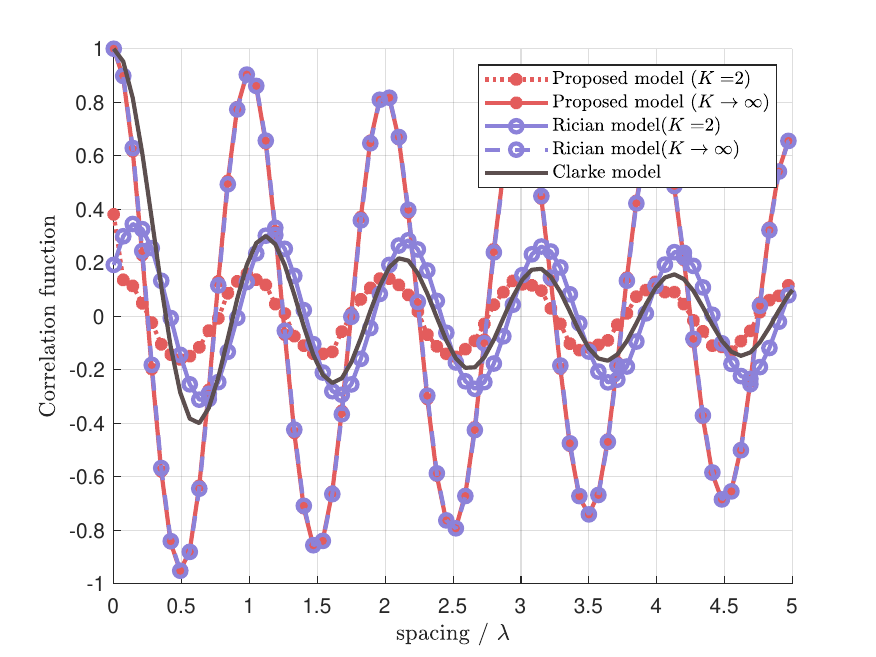}  
    \end{minipage} 
}%
		\caption{The channel correlation function vs. antenna spacing in transmitter at transceiver distance (a) $d_{\mathrm{RT}}=0.4\lambda$ and (b) $d_{\mathrm{RT}}=10\lambda$. }    
		\label{fig:Cor_KFactor}  
		\vspace{-0.5cm}
	\end{figure*}
In this section, we provide the numerical simulation of the channel model, i.e., channel correlation, eigenvalue distribution, and channel capacity, in the multi-path environment. In addition, the theoretical channel capacity and capacity region of MISO systems are also evaluated. 
\subsection{Channel Modeling}
The effectiveness of the proposed channel model is verified in this part, and the baselines (Clarke's model and Rician channel) are provided for comparison. The operating frequency is $5$ GHz.  
\subsubsection{Correlation of the proposed model vs baselines}
We compare the correlation function in terms of spacing in the transmitter of the proposed channel model and benchmarks (Clarke model and Rician model) with different $K$-factors at transceiver distance $d_{\mathrm{RT}}=0.4\lambda$ and $d_{\mathrm{RT}}=10\lambda$ in Fig.~\ref{fig:Cor_KFactor}. The parameter setting is $N_s=100$ and $N_r=1$.  

As shown in Fig.~\ref{fig:Cor_KFactor} (a), a closer transceiver distance has more dominant LoS components, thus the channel correlation of a small $K$-factor (e.g., $K=2$) easily matches with the $K\rightarrow\infty$ scenario. Moreover, Fig.~\ref{fig:Cor_KFactor} (a) confirms the effectiveness of our proposed channel model, aligning well with the Clarke model. However, discrepancies emerge for larger transceivers, as shown in Fig.~\ref{fig:Cor_KFactor} (b), where a gap appears between the proposed model and two benchmarks for $K=2$. This reveals limitations in the conventional Rician and Clarke models to accurately describe channel correlation behavior in densely spaced antenna scenarios across various multi-path conditions. 

 \subsubsection{Eigenvalues of the polarized channel}  
\begin{figure*}[htbp] 
		\centering  
		\subfigure[Proposed channel ($K=10$)]{
    \begin{minipage}[t]{0.45\linewidth}
        \centering
        \includegraphics[width=1\linewidth]{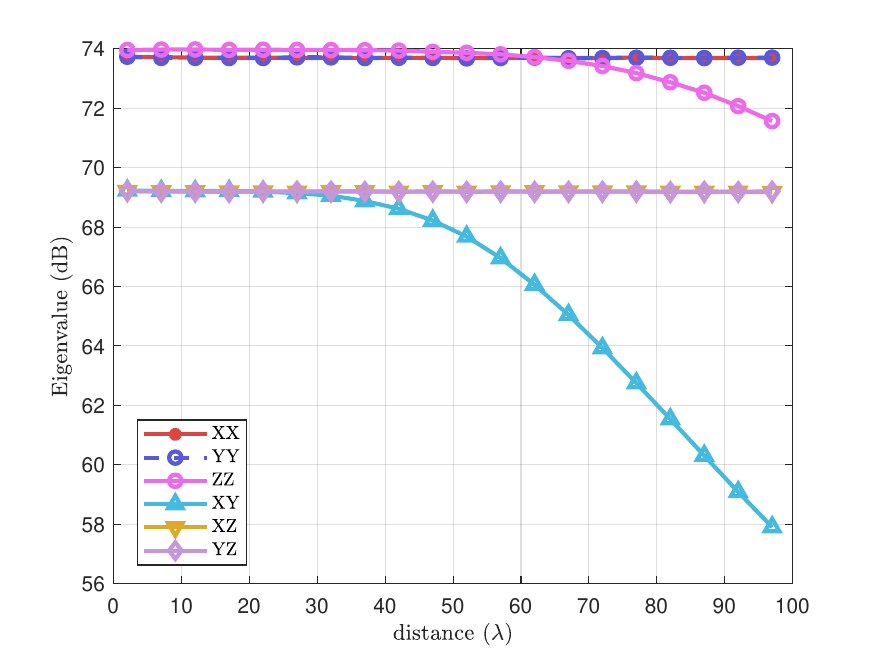}  
        \vspace{0.02cm}  
    \end{minipage}%
} 
\subfigure[Proposed channel ($K\rightarrow \infty$)]{
    \begin{minipage}[t]{0.45\linewidth}
        \centering
        \includegraphics[width=1\linewidth]{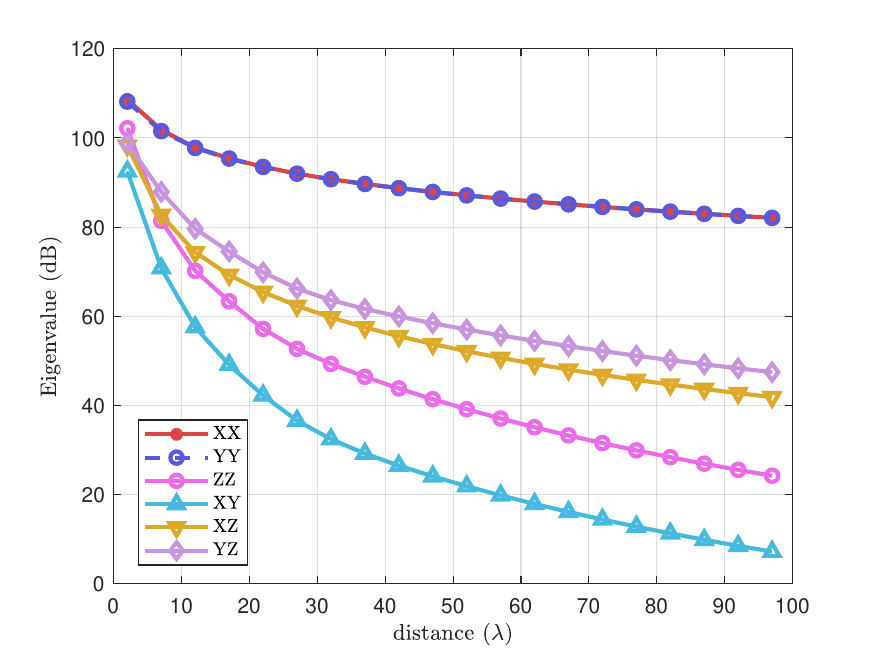}  
    \end{minipage} 
}%
		\caption{The eigenvalues of the proposed polarized channel ($N_s=36$ and $N_r=16$) for different $K$-factors at varying transmitter-receiver distances from $2\lambda$ to $100\lambda$.}    
		\label{fig:eigenvalues}  
		\vspace{-0.5cm}
	\end{figure*}

We compare eigenvalues of co-polarization and cross-polarization channels for $K=10$ and $K\rightarrow \infty$ across transceiver distances from $d_{\mathrm{RT}}=2\lambda$ to $100\lambda$ in Fig.~\ref{fig:eigenvalues}.  The parameter setting is $N_s=36$, $N_r=16$, and the antenna spacing is $0.4\lambda$.
 
 In the multi-path environment that involves rich NLoS paths, as shown in Fig.~\ref{fig:eigenvalues} (a), co-polarized components (XX, YY, and ZZ) exhibit larger eigenvalues than cross-polarized components (XY, XZ, and YZ). Most polarization components show similar eigenvalue behavior due to the rich randomness, except ZZ and XY which decrease notably in the far-field region due to fast-decaying EM waves. Contrarily, in Fig.~\ref{fig:eigenvalues} (b), a LoS-dominant scenario with larger $K$-factors, all polarization components consistently decrease with distance, with the polarization component parallel to the propagation direction (ZZ) diminishing faster compared to XX and YY. Cross-polarization components also become more prominent, emphasizing the importance of cross-polarization interference elimination techniques in wireless systems, particularly in near-field communications.

\subsubsection{Capacity of the channel with/without practical constraints} 
\begin{figure}  
	\begin{center}
		{\includegraphics[width=0.45 \textwidth]{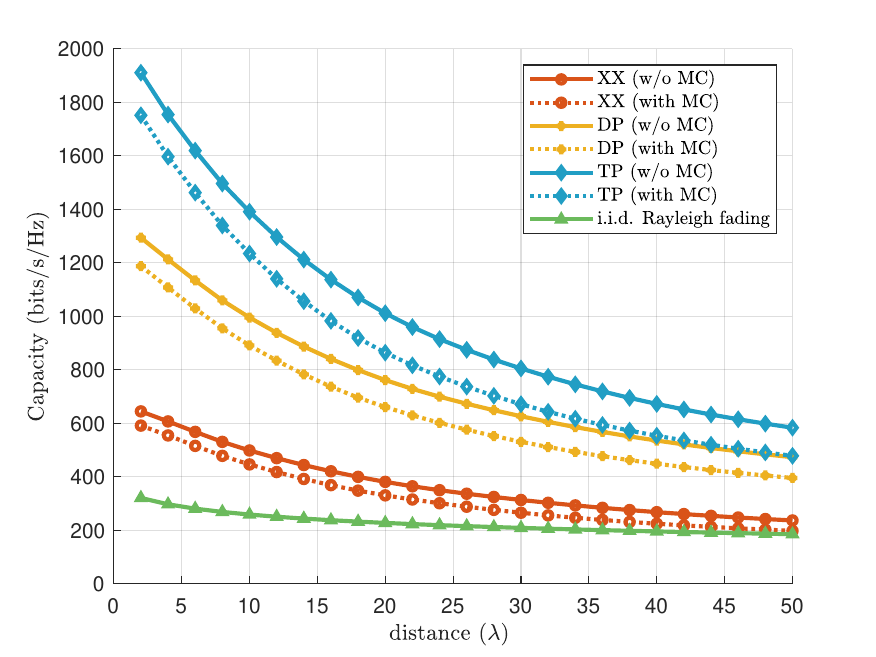}}  
		\caption{The capacity of the proposed channel ($N_s=36$ and $N_r=16$) with/without practical constraints and i.i.d. Rayleigh fading model at varying distances from $2\lambda$ to $50\lambda$. }  
		\label{fig:CapacityWOMC} 
	\end{center}
\end{figure}

We compare the capacity of the LoS channel with practical constraints that contain mutual coupling matrix and reduced energy efficiency (denoted as ``with MC") in Fig.~\ref{fig:CapacityWOMC}. For comparison, the polarized channel without practical constraints (denoted as ``w/o MC") and i.i.d. Rayleigh fading channel at varying distances from $2\lambda$ to $50\lambda$ are also presented.  Since the impacts of $K$-factor have been evaluated in the former simulations, the multi-path channel is constructed using \eqref{equ:distriChannel} without specifying $K$-factors in \eqref{equ:KComposedChannel}, and the cavity volume is $(400\lambda)^3$.  The parameter setting is $N_s=36$, $N_r=16$, and the antenna spacing is $0.4\lambda$.

First, the capacity of the practical case is lower than that of the ideal case due to the coupling effects and reduced energy efficiency, and such a gap is more evident in dual-polarization (DP) and tri-polarization (TP) cases. Then, as the distance goes further, e.g., from $d_{\mathrm{RT}}=2\lambda$ to $d_{\mathrm{RT}}=50\lambda$, the capacity of channels for all polarization states decreases and the gap between TP and DP cases narrows as well, which is partly attributed to the vanishing polarization component and the weaker channel gains. Additionally, the gap between the conventional i.i.d. Rayleigh fading channel and the proposed channel occurs in the far-field region (e.g., $d_{\mathrm{RT}}=50\lambda$). This confirms that the conventional Rayleigh fading model underestimates capacity in far-field regions, with even greater underestimation in near-field regions. 
 
\subsubsection{Capacity of the multi-path and LoS channels} 
\begin{figure}  
	\begin{center}
		{\includegraphics[width=0.45 \textwidth]{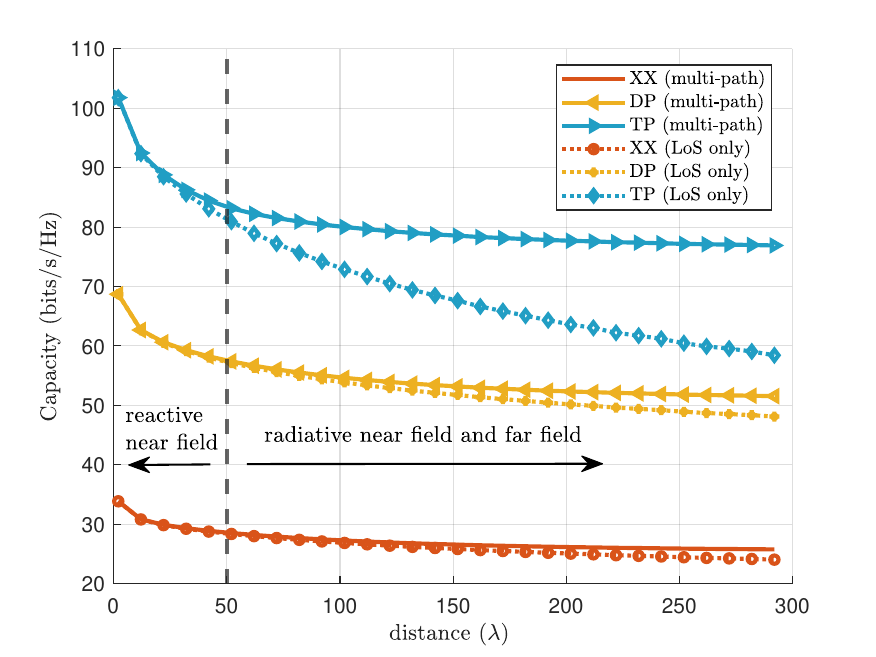}}  
		\caption{The capacity of the proposed channel model in multi-path and LoS-only cases at varying transceiver distances from $2\lambda$ to $300\lambda$. }  
		\label{fig:MP_LOS_100_1} 
	\end{center}
\end{figure}

The comparison of the polarized channel in multi-path and LoS-only scenarios at varying distances from $2\lambda$ to $300\lambda$ is given in Fig.~\ref{fig:MP_LOS_100_1}. The parameter setting is $N_s=100$ and $N_r=1$. It can be observed that the proposed channel model effectively describes both near-field gain and multi-path effects in the radiative near field, showing advantages over the free-space Green's functions in depicting radiative near-field communications. At the same time, in the reactive near-field region, both the composite and LoS channels exhibit similar capacity performance due to negligible multi-path effects overshadowed by dominant LoS components. However, in the radiative near-field and far-field regions, the multi-path channel shows significantly improved performance attributed to backscattering effects. This highlights near-field gains in near-field communications and multi-path dominance in far-field scenarios, and the capacity enhancement is more pronounced in tri-polarized (TP) channels due to richer polarized components.

\subsection{Theoretical capacity evaluation}

\subsubsection{Theoretical capacity bound} 
\begin{figure}  
	\begin{center}
		{\includegraphics[width=0.48 \textwidth]{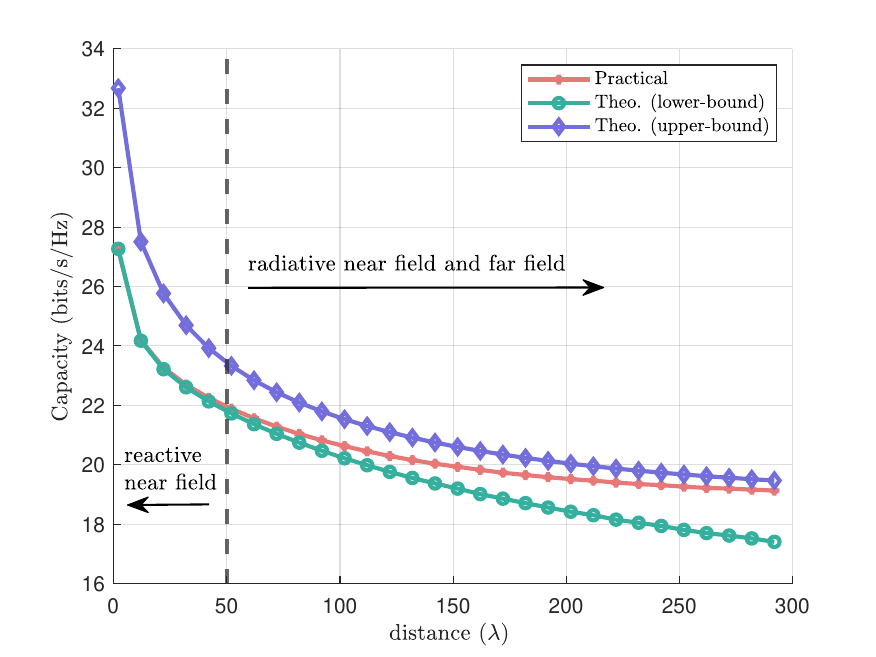}}  
		\caption{The comparison of the practical channel capacity and theoretical capacity at varying transceiver distances from $2\lambda$ to $300\lambda$. }  
		\label{fig:theo_100_1} 
	\end{center}
\end{figure}

The theoretical bound of the proposed channel model is given in Fig.~\ref{fig:theo_100_1}. The parameter setting is $N_s=100$ and $N_r=1$. It can be observed that the theoretical lower bound perfectly matches the practical channel model in the reactive near-field region and the theoretical upper bound predicts the channel capacity in the radiative near-field and far-field zones, thus validating the effectiveness of the theoretical analysis. 

 \subsubsection{Capacity region of the two-user channel} 
\begin{figure}  
	\begin{center}
		{\includegraphics[width=0.48 \textwidth]{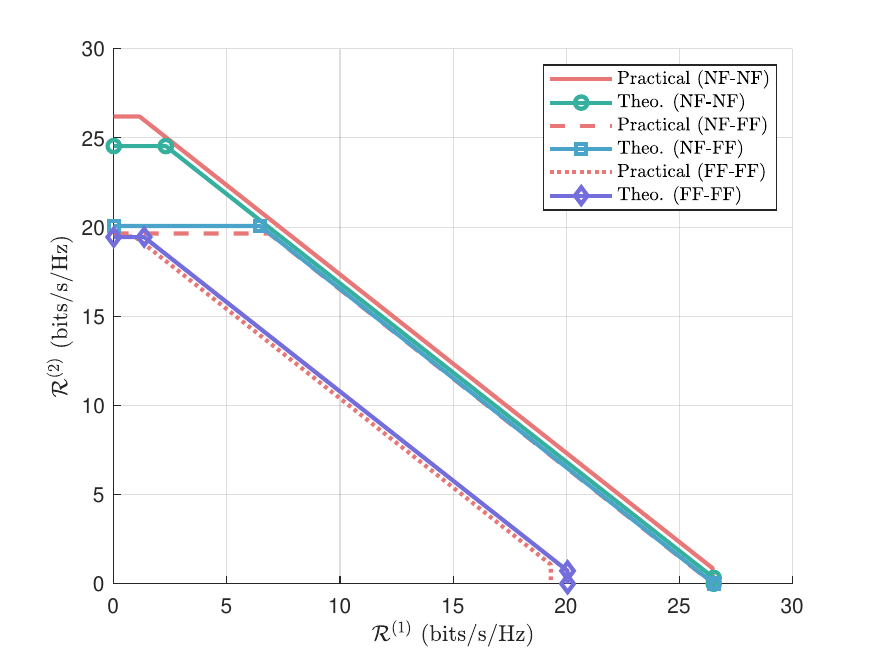}}  
		\caption{The practical and theoretical capacity region of two-user systems in different communication zones. }  
		\label{fig:CapacityRegion} 
	\end{center}
\end{figure}

We compare the practical and theoretical capacity region of the two-user channel for different communication regions in Fig.~\ref{fig:CapacityRegion}. In the parameter setting, two single-antenna users communicate with the transmitter equipped with $N_s=100$ antennas, and the locations of two users are mainly classified in three cases: 1) NF-NF: two near-field users located at $d_{\mathrm{RT}}=3\lambda$ and $4\lambda$, respectively; 2) NF-FF: one near-field user located at $d_{\mathrm{RT}}=3\lambda$ and the other far-field user located at $200\lambda$; and 3) FF-FF: two far-field users located at $d_{\mathrm{RT}}=200\lambda$ and $300\lambda$, respectively.  

 It can be observed that the theoretical analysis predicts the capacity region across all communication regions, affirming the robustness of the performance analysis. Additionally, greater separation between two users reduces inter-user interference, as shown in Fig.~\ref{fig:CapacityRegion}. For instance, in the NF-FF case where users are located at $d_{\mathrm{RT}}=3\lambda$ and $200\lambda$, the capacity of user 1, i.e., $\mathcal{R}^{(1)}$, is higher than that in other two cases in the presence of complete inter-user interference. 

\begin{figure}[!htbp]\quad
		\centering  
		\subfigure[Two near-field users]{
    \begin{minipage}[t]{1\linewidth}
        \centering
        \includegraphics[width=1\linewidth]{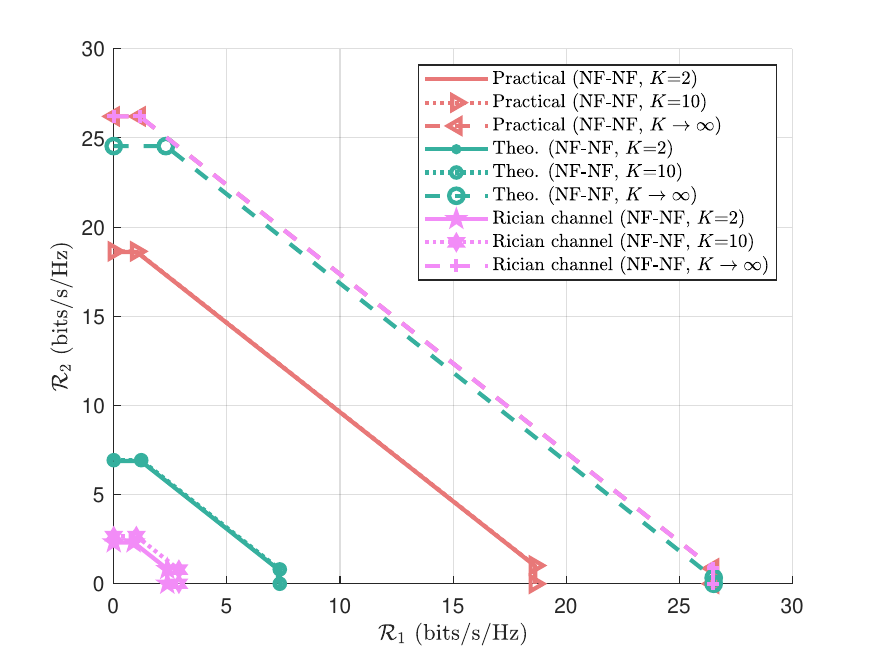}
    \end{minipage}%
}%
\quad 
\subfigure[One near-field user and one far-field user]{
    \begin{minipage}[t]{1\linewidth}
        \centering
        \includegraphics[width=1\linewidth]{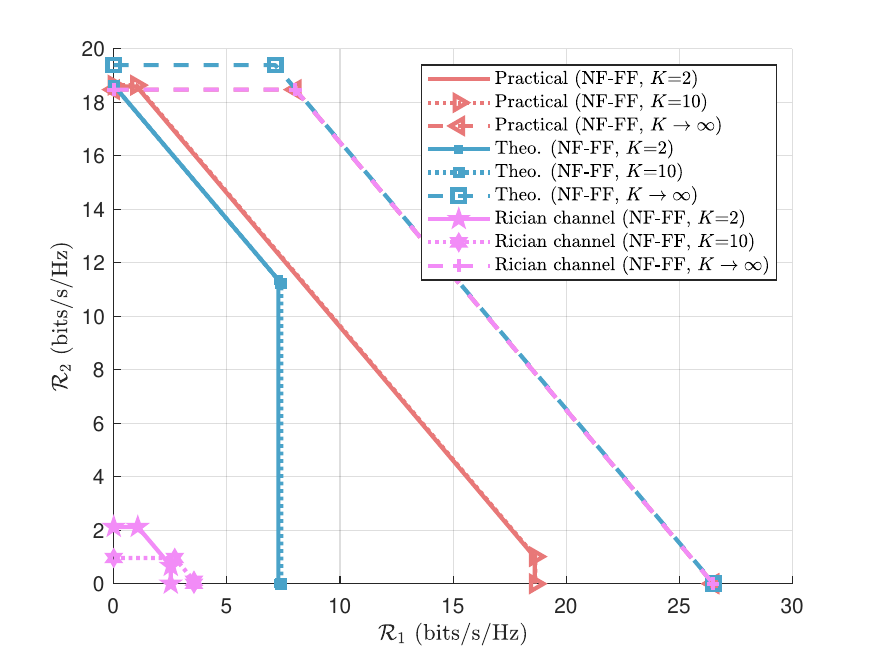}  
    \end{minipage} 
} 
\quad
\subfigure[Two far-field users]{
    \begin{minipage}[t]{1\linewidth}
        \centering
        \includegraphics[width=1\linewidth]{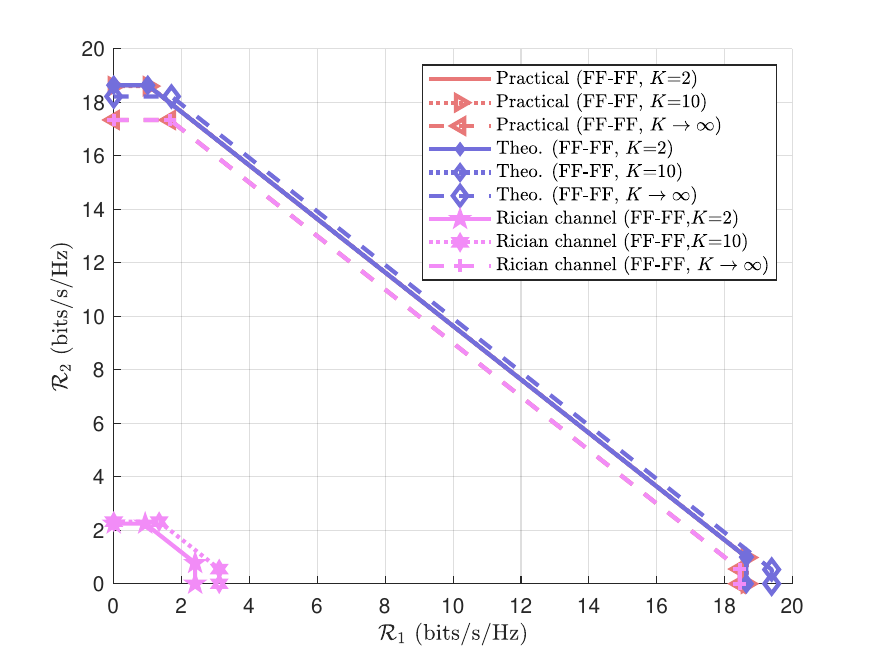}  
    \end{minipage} 
} 
		\caption{The capacity region of two-user channels with different $K$-factors.}    
		\label{fig:CapRegionKfactor}  
		\vspace{-0.5cm}
	\end{figure}

The capacity region comparison of the proposed channel model and Rician channel with different $K$-factors for two users are given in Fig.~\ref{fig:CapRegionKfactor}, simulating three cases from Fig.~\ref{fig:CapacityRegion}. In the NF-NF case (Fig.~\ref{fig:CapacityRegion} (a)),  the traditional Rician channels with $K=2$ and $K=10$ underestimate the capacity region of practical systems, aligning with the LoS-only scenario due to an imperfect description of scattered EM waves in near-field regions. This underestimation also occurs in both NF-FF (Fig.~\ref{fig:CapacityRegion} (b)) and FF-FF cases (Fig.~\ref{fig:CapacityRegion} (c)), supporting the observation in  Fig.~\ref{fig:CapacityWOMC}, further demonstrating that conventional Rician and Rayleigh fading channel models cannot accurately describe the capacity performance. Additionally, increasing the $K$-factor can effectively enlarge the capacity region in NF-NF and NF-FF cases due to the increasing ratio of dominant LoS components in near-field regions. Such a gain disappears in the far-field case, as shown in the FF-FF case (Fig.~\ref{fig:CapacityRegion} (c)) due to the vanishing evanescent waves in the far-field region. 

\section{Conclusions} \label{sec:conclusion} 
This paper presented a universal EM-compliant channel model for HMIMO communication systems based on stochastic Green's functions. By incorporating a large number of planar waves, the proposed channel can be statistically represented in a probabilistic framework. The asymptotic channel capacity and capacity region are derived based on the physics-oriented channel models.  Our simulation results substantiated the effectiveness of the proposed channel model, which can accurately describe both spatial correlation and polarization effects. It can be observed that the traditional Rician cannot accurately describe the channel correlation, and conventional i.i.d. Rayleigh systems underestimate the single-polarized channel in near-/middle-field regions.  In addition, the proposed channel model describes the radiative near-field region more accurately than free-space Green's functions. At the same time, in the reactive near-field region, both composite and LoS channels show similar capacity performance due to negligible multi-path effects dominated by LoS components. However, the multi-path effects improve the radiative near-field/far-field channel capacity even in the absence of near-field effects. These capacity gains are more pronounced in tri-polarized channels compared to dual-polarized and single-polarized channels. Furthermore, the efficient theoretical analysis substantiates the effectiveness of the proposed channel model and the incapability of the traditional models in describing the system capacity and two-user capacity region.
 
However, the proposed probabilistic model still encounters significant challenges, primarily due to the difficulty in acquiring well-defined parameters such as volume and shape in diverse scenarios. Extending the model to account for these parameters across varying wireless environments remains a complex and demanding task, offering valuable opportunities for further research and exploration.

 \color{black}
 
 \section*{Appendix}
 The variances of $D_{xy}$  and  $D_{yx}$ are  
 \begin{equation}\notag
    \begin{aligned}
        \mathbb{E} &[D_{xy}^2] = \mathbb{E}[D_{yx}^2]   = \mathbb{E}[(\mathbf{\Psi}_i^x(\mathbf{r}), \mathbf{\Psi}_i^y(\mathbf{r}'))^2]  \\  
       =& \frac{1}{8\pi^2V^2}\mathbb{E} \left[ \left( \frac{ 1}{4} \sin^2 (2\phi_n) \cos^4 \psi_n  \right. \right.    - 2 \sin\psi_n \cos^3 \psi_n  \\
       & \cdot \cos  \phi_n  \sin^3 \phi_n \cos \theta_n    + \frac{1}{4} \sin^2 (2\psi_n) \sin^4 \phi_n \cos^2 \theta_n   \\
       & + 2 \sin\psi_n \cos^3 \psi_n \sin  \phi_n  \cos^3 \phi_n \cos \theta_n   -\frac{1}{4} \sin^2  (2\psi_n)  \\
       \quad & \cdot \sin^2 (2\phi_n) \cos^2 \theta_n  + 2 \sin^3  \psi_n \cos \psi_n \cos \phi_n \sin^3 \phi_n \cos^3 \theta_n \\
       & + \frac{1}{4} \sin^2  (2\psi_n) \cos^4   \phi_n  \cos^2 \theta_n  - 2\cos\psi_n \sin^3   \psi_n  \\
       & \cdot \sin  \phi_n  \cos^3 \phi_n \cos^3\theta_n \left. +\frac{1}{4} \sin^4   \psi_n  \sin^2 (2\phi_n) \cos^4 \theta_n \right) \\
       & \left.  \cos^2 (k_i R \cos \theta) \right] \\
        =&  \frac{1}{8\pi^2V^2} \left[   \frac{ \pi^2}{16} \left( \frac{2}{3}+2\frac{\sin \left(2 k_i R\right)}{2 k_i R}+4\frac{ \cos \left(2 k_i R\right)}{\left(2 k_i R\right)^2}   \right. \right.  \left. -4\frac{  \sin \left(2 k_i R\right)}{\left(2 k_i R\right)^3}  \right) \\
        & +   \frac{ \pi^2}{ 16} \cdot \left( \frac{18}{5} +6 \frac{\sin(2k_i R)}{2k_i R}   \right.   + 12\frac{\cos(2k_i R)}{(2k_i R)^2}   - 36\frac{\sin(2k_i R)}{(2k_i R)^3} \\ 
        & \left. \left.  -  72\frac{ \cos(2k_i R)}{(2k_i R)^4}   + 72\frac{ \sin(2k_i R)}{(2k_i R)^5} \right) \right]\\
        =& \frac{1}{128 V^2} \left[ \frac{64}{15} +8 \frac{\sin(2k_i R)}{2k_i R}  + 16\frac{\cos(2k_i R)}{(2k_i R)^2} \right.   \\
        & \left.  - 40\frac{\sin(2k_i R)}{(2k_i R)^3} -  72\frac{ \cos(2k_i R)}{(2k_i R)^4}  + 72\frac{ \sin(2k_i R)}{(2k_i R)^5}   \right].
    \end{aligned}
\end{equation}

\section*{Acknowledgment}
The authors would like to express their gratitude to Wei E. I. Sha for the valuable and insightful discussions, which significantly enhanced the scope and clarity of this paper.

\bibliographystyle{IEEEtran.bst}
\bibliography{capacityHMIMO}

\begin{thebibliography}{10}
\providecommand{\url}[1]{#1}
\csname url@samestyle\endcsname
\providecommand{\newblock}{\relax}
\providecommand{\bibinfo}[2]{#2}
\providecommand{\BIBentrySTDinterwordspacing}{\spaceskip=0pt\relax}
\providecommand{\BIBentryALTinterwordstretchfactor}{4}
\providecommand{\BIBentryALTinterwordspacing}{\spaceskip=\fontdimen2\font plus
\BIBentryALTinterwordstretchfactor\fontdimen3\font minus
  \fontdimen4\font\relax}
\providecommand{\BIBforeignlanguage}[2]{{%
\expandafter\ifx\csname l@#1\endcsname\relax
\typeout{** WARNING: IEEEtran.bst: No hyphenation pattern has been}%
\typeout{** loaded for the language `#1'. Using the pattern for}%
\typeout{** the default language instead.}%
\else
\language=\csname l@#1\endcsname
\fi
#2}}
\providecommand{\BIBdecl}{\relax}
\BIBdecl

\bibitem{5595728}
T.~L. Marzetta, ``Noncooperative cellular wireless with unlimited numbers of
  base station antennas,'' \emph{IEEE Trans. Wirel. Commun.}, vol.~9, no.~11,
  pp. 3590--3600, Nov. 2010.

\bibitem{shlezinger2020dynamic_all}
N.~Shlezinger, G.~C. Alexandropoulos, M.~F. Imani, Y.~C. Eldar, and D.~R.
  Smith, ``Dynamic metasurface antennas for {6G} extreme massive {MIMO}
  communications,'' \emph{IEEE Wireless Commun.}, vol.~28, no.~2, pp. 106--113,
  Jan. 2021.

\bibitem{9475156}
D.~Dardari and N.~Decarli, ``Holographic communication using intelligent
  surfaces,'' \emph{IEEE Commun. Mag.}, vol.~59, no.~6, pp. 35--41, Jun. 2021.

\bibitem{10505154}
J.~Xu, L.~You, G.~C. Alexandropoulos, X.~Yi, W.~Wang, and X.~Gao, ``Near-field
  wideband extremely large-scale {MIMO} transmissions with holographic
  metasurface-based antenna arrays,'' \emph{IEEE Trans. Wirel. Commun.}, pp.
  1--1, 2024.

\bibitem{10542433}
S.~S.~A. Yuan, J.~Wu, H.~Xu, T.~Wang, D.~Li, X.~Chen, C.~Huang, S.~Sun,
  S.~Zheng, X.~Zhang, E.~Li, and W.~E.~I. Sha, ``Breaking the
  degrees-of-freedom limit of holographic {MIMO} communications: A {3-D}
  antenna array topology,'' \emph{IEEE Trans. Veh. Technol.}, pp. 1--13, 2024.

\bibitem{JINDAN_CS}
J.~Xu, C.~Yuen, C.~Huang, N.~Ul~Hassan, G.~C. Alexandropoulos, M.~Di~Renzo, and
  M.~Debbah, ``Reconfiguring wireless environments via intelligent surfaces for
  6g: Reflection, modulation, and security,'' \emph{Sci. China Inf. Sci.},
  vol.~66, no.~3, p. 130304, 2023.

\bibitem{ramezani2023near}
P.~Ramezani and E.~Bj{\"o}rnson, ``Near-field beamforming and multiplexing
  using extremely large aperture arrays,'' in \emph{Fundamentals of 6G
  Communications and Networking}.\hskip 1em plus 0.5em minus 0.4em\relax
  Springer, 2023, pp. 317--349.

\bibitem{9136592}
C.~Huang, S.~Hu, G.~C. Alexandropoulos, A.~Zappone, C.~Yuen, R.~Zhang, M.~D.
  Renzo, and M.~Debbah, ``Holographic {MIMO} surfaces for {6G} wireless
  networks: Opportunities, challenges, and trends,'' \emph{IEEE Wirel.
  Commun.}, vol.~27, no.~5, pp. 118--125, Oct. 2020.

\bibitem{RISE6G_COMMAG}
E.~Calvanese~Strinati, G.~C. Alexandropoulos, H.~Wymeersch, B.~Denis,
  V.~Sciancalepore, R.~D'Errico, A.~Clemente, D.-T. Phan-Huy, E.~De~Carvalho,
  and P.~Popovski, ``Reconfigurable, intelligent, and sustainable wireless
  environments for {6G} smart connectivity,'' \emph{IEEE Commun. Mag.},
  vol.~59, no.~10, pp. 99--105, Oct. 2021.

\bibitem{9779586}
L.~Wei, C.~Huang, G.~C. Alexandropoulos, W.~E.~I. Sha, Z.~Zhang, M.~Debbah, and
  C.~Yuen, ``Multi-user holographic {MIMO} surfaces: Channel modeling and
  spectral efficiency analysis,'' \emph{IEEE J. Sel. Top. Signal Process.},
  vol.~16, no.~5, pp. 1112--1124, Aug. 2022.

\bibitem{8741198}
C.~Huang, A.~Zappone, G.~C. Alexandropoulos, M.~Debbah, and C.~Yuen,
  ``Reconfigurable intelligent surfaces for energy efficiency in wireless
  communication,'' \emph{IEEE Trans. Wirel. Commun.}, vol.~18, no.~8, pp.
  4157--4170, Aug. 2019.

\bibitem{10663346}
L.~Jin, X.~Xu, S.~Han, X.~Chi, P.~Zhang, and C.~Yuen, ``Achievable rate of
  linear holographic {MIMO} with arbitrary aperture-length,'' \emph{IEEE Trans.
  Wirel. Commun.}, vol.~23, no.~11, pp. 16\,742--16\,756, Nov. 2024.

\bibitem{10130641}
J.~An, C.~Yuen, C.~Huang, M.~Debbah, H.~Vincent~Poor, and L.~Hanzo, ``A
  tutorial on holographic {MIMO} communications—{Part I}: Channel modeling
  and channel estimation,'' \emph{IEEE Commun. Lett.}, vol.~27, no.~7, pp.
  1664--1668, Jul. 2023.

\bibitem{bjornson2019massive}
E.~Bj{\"o}rnson, L.~Sanguinetti, H.~Wymeersch, J.~Hoydis, and T.~L. Marzetta,
  ``Massive {MIMO} is a reality—what is next?: Five promising research
  directions for antenna arrays,'' \emph{Digital Signal Processing}, vol.~94,
  pp. 3--20, 2019.

\bibitem{10232975}
T.~Gong, P.~Gavriilidis, R.~Ji, C.~Huang, G.~C. Alexandropoulos, L.~Wei,
  Z.~Zhang, M.~Debbah, H.~V. Poor, and C.~Yuen, ``Holographic {MIMO}
  communications: Theoretical foundations, enabling technologies, and future
  directions,'' \emph{IEEE Commun. Surv. Tutor.}, vol.~26, no.~1, pp. 196--257,
  Firstquarter 2024.

\bibitem{9530717}
S.~Basharat, S.~Hassan, H.~Pervaiz, A.~Mahmood, Z.~Ding, and M.~Gidlund,
  ``Reconfigurable intelligent surfaces: Potentials, applications, and
  challenges for {6G} wireless networks,'' \emph{IEEE Wirel. Commun.}, pp.
  1--8, 2021.

\bibitem{FranceschettiCapacityWirelessNetworks2009}
M.~Franceschetti, M.~D. Migliore, and P.~Minero, ``The capacity of wireless
  networks: Information-theoretic and physical limits,'' \emph{IEEE Trans.
  Inform. Theory}, vol.~55, no.~8, pp. 3413--3424, Aug. 2009.

\bibitem{FranceschettiInformationCarriedScattered2015}
M.~Franceschetti, M.~D. Migliore, P.~Minero, and F.~Schettino, ``The
  information carried by scattered waves: Near-field and nonasymptotic
  regimes,'' \emph{IEEE Trans. Antennas Propagat.}, vol.~63, no.~7, pp.
  3144--3157, Jul. 2015.

\bibitem{9495942}
A.~S. de~Sena, P.~H.~J. Nardelli, D.~B.~d. Costa, U.~S. Dias, P.~Popovski, and
  C.~B. Papadias, ``Dual-polarized {IRSs} in uplink {MIMO-NOMA} networks: An
  interference mitigation approach,'' \emph{IEEE Wirel. Commun. Lett.},
  vol.~10, no.~10, pp. 2284--2288, 2021.

\bibitem{DENGZhiji41}
J.~SHEN, J.~YANG, C.~ZHU, Z.~DENG, and C.~HUANG, ``Near-field beam training for
  holographic {MIMO} communications: Typical methods, challenges and future
  directions,'' \emph{ZTE Communications}, vol.~22, no.~1, pp. 41--52, 2024.

\bibitem{9497725}
Y.~Han, X.~Li, W.~Tang, S.~Jin, Q.~Cheng, and T.~J. Cui, ``Dual-polarized
  {RIS}-assisted mobile communications,'' \emph{IEEE Trans. Wirel. Commun.},
  vol.~21, no.~1, pp. 591--606, Jan. 2022.

\bibitem{10443720}
L.~Xu, L.~Cheng, N.~Wong, Y.-C. Wu, and H.~V. Poor, ``Overcoming beam squint in
  mmwave {MIMO} channel estimation: A {Bayesian} multi-band sparsity
  approach,'' \emph{IEEE Trans. Signal Process.}, vol.~72, pp. 1219--1234,
  2024.

\bibitem{4447351}
M.~A. Jensen and J.~W. Wallace, ``Capacity of the continuous-space
  electromagnetic channel,'' \emph{IEEE Trans. Antennas Propag.}, vol.~56,
  no.~2, pp. 524--531, Feb. 2008.

\bibitem{7805306}
T.~J. Lim and M.~Franceschetti, ``Information without rolling dice,''
  \emph{IEEE Trans. Inf. Theory}, vol.~63, no.~3, pp. 1349--1363, Mar. 2017.

\bibitem{9650519}
S.~S.~A. Yuan, Z.~He, X.~Chen, C.~Huang, and W.~E.~I. Sha, ``Electromagnetic
  effective degree of freedom of an {MIMO} system in free space,'' \emph{IEEE
  Antennas Wirel. Propag. Lett.}, vol.~21, no.~3, pp. 446--450, Mar. 2022.

\bibitem{8024037}
B.~Jonsson, S.~Shi, L.~Wang, F.~Ferrero, and L.~Lizzi, ``On methods to
  determine bounds on the {Q}-factor for a given directivity,'' \emph{IEEE
  Trans. Antennas Propag.}, vol.~65, no.~11, pp. 5686--5696, Nov. 2017.

\bibitem{1417209}
A.~Yaghjian and S.~Best, ``Impedance, bandwidth, and {Q} of antennas,''
  \emph{IEEE Trans. Antennas Propag.}, vol.~53, no.~4, pp. 1298--1324, Apr.
  2005.

\bibitem{9399839}
C.~Ehrenborg, M.~Gustafsson, and M.~Capek, ``Capacity bounds and degrees of
  freedom for {MIMO} antennas constrained by {Q}-factor,'' \emph{IEEE Trans.
  Antennas Propag.}, vol.~69, no.~9, pp. 5388--5400, Sept. 2021.

\bibitem{8998551}
C.~Ehrenborg and M.~Gustafsson, ``Physical bounds and radiation modes for
  {MIMO} antennas,'' \emph{IEEE Trans. Antennas Propag.}, vol.~68, no.~6, pp.
  4302--4311, Jun. 2020.

\bibitem{10628002}
R.~Ji, C.~Huang, X.~Chen, W.~E.~I. Sha, L.~Dai, J.~He, Z.~Zhang, C.~Yuen, and
  M.~Debbah, ``Electromagnetic hybrid beamforming for holographic {MIMO}
  communications,'' \emph{IEEE Trans. Wirel. Commun.}, vol.~23, no.~11, pp.
  15\,973--15\,986, Nov. 2024.

\bibitem{10103817}
L.~Wei, C.~Huang, G.~C. Alexandropoulos, Z.~Yang, J.~Yang, W.~E.~I. Sha,
  Z.~Zhang, M.~Debbah, and C.~Yuen, ``Tri-polarized holographic {MIMO} surfaces
  for near-field communications: Channel modeling and precoding design,''
  \emph{IEEE Trans. Wirel. Commun.}, vol.~22, no.~12, pp. 8828--8842, Dec.
  2023.

\bibitem{9724113}
A.~Pizzo, L.~Sanguinetti, and T.~L. Marzetta, ``Fourier plane-wave series
  expansion for holographic {MIMO} communications,'' \emph{IEEE Trans. Wirel.
  Commun.}, pp. 1--16, Mar. 2022.

\bibitem{10018012}
S.~Lin, S.~Luo, S.~Ma, J.~Feng, Y.~Shao, Z.~B. Drikas, B.~D. Addissie, S.~M.
  Anlage, T.~Antonsen, and Z.~Peng, ``Predicting statistical wave physics in
  complex enclosures: A stochastic dyadic {Green's} function approach,''
  \emph{IEEE Trans. Electromagn. Compat.}, vol.~65, no.~2, pp. 436--453, Apr.
  2023.

\bibitem{10381502}
S.~Yu, W.~Chen, and H.~V. Poor, ``Real-time monitoring of chaotic systems with
  known dynamical equations,'' \emph{IEEE Trans. Signal Process.}, vol.~72, pp.
  1251--1268, 2024.

\bibitem{8952896}
S.~Lin, Z.~Peng, and T.~M. Antonsen, ``A stochastic {Green’s} function for
  solution of wave propagation in wave-chaotic environments,'' \emph{IEEE
  Trans. Antennas Propag.}, vol.~68, no.~5, pp. 3919--3933, May 2020.

\bibitem{arnoldus2001representation}
H.~F. Arnoldus, ``Representation of the near-field, middle-field, and far-field
  electromagnetic {Green’s} functions in reciprocal space,'' \emph{JOSA B},
  vol.~18, no.~4, pp. 547--555, Apr. 2001.

\bibitem{6557073}
C.~Lemoine, E.~Amador, P.~Besnier, J.-M. Floc'h, and A.~Laisné, ``Antenna
  directivity measurement in reverberation chamber from rician {K}-factor
  estimation,'' \emph{IEEE Trans. Antennas Propag.}, vol.~61, no.~10, pp.
  5307--5310, Oct. 2013.

\bibitem{4012433}
C.~L. Holloway, D.~A. Hill, J.~M. Ladbury, P.~F. Wilson, G.~Koepke, and
  J.~Coder, ``On the use of reverberation chambers to simulate a {Rician} radio
  environment for the testing of wireless devices,'' \emph{IEEE Trans. Antennas
  Propag.}, vol.~54, no.~11, pp. 3167--3177, Nov. 2006.

\bibitem{1167256}
R.~Janaswamy, ``Effect of element mutual coupling on the capacity of fixed
  length linear arrays,'' \emph{IEEE Antennas Wirel. Propag. Lett.}, vol.~1,
  pp. 157--160, 2002.

\bibitem{1291781}
S.~Durrani and M.~Bialkowski, ``Effect of mutual coupling on the interference
  rejection capabilities of linear and circular arrays in cdma systems,''
  \emph{IEEE Trans. Antennas Propag.}, vol.~52, no.~4, pp. 1130--1134, Apr.
  2004.

\bibitem{9512430}
M.~A. Azam, A.~K. Dutta, and A.~Mukherjee, ``Performance analysis of dipole
  antenna based planar arrays with mutual coupling and antenna position error
  in {mmWave} hybrid system,'' \emph{IEEE Trans. Veh. Technol.}, vol.~70,
  no.~10, pp. 10\,209--10\,221, Oct. 2021.

\bibitem{kildal2015foundations}
P.-S. Kildal, \emph{Foundations of antenna engineering: a unified approach for
  line-of-sight and multipath}.\hskip 1em plus 0.5em minus 0.4em\relax Artech
  House, 2015.

\bibitem{299601}
A.~Roscoe and R.~Perrott, ``Large finite array analysis using infinite array
  data,'' \emph{IEEE Trans. Antennas Propag.}, vol.~42, no.~7, pp. 983--992,
  Jul. 1994.

\bibitem{8718501}
M.~Gustafsson and M.~Capek, ``Maximum gain, effective area, and directivity,''
  \emph{IEEE Trans. Antennas Propag.}, vol.~67, no.~8, pp. 5282--5293, Aug.
  2019.

\bibitem{1237143}
S.~Vishwanath, N.~Jindal, and A.~Goldsmith, ``Duality, achievable rates, and
  sum-rate capacity of {Gaussian MIMO} broadcast channels,'' \emph{IEEE Trans.
  Inf. Theory}, vol.~49, no.~10, pp. 2658--2668, Oct. 2003.

\bibitem{850678}
I.~Telatar and D.~Tse, ``Capacity and mutual information of wideband multipath
  fading channels,'' \emph{IEEE Trans. Inf. Theory}, vol.~46, no.~4, pp.
  1384--1400, Jul. 2000.

\bibitem{7024192}
K.-H. Chen and J.-F. Kiang, ``Effect of mutual coupling on the channel capacity
  of {MIMO} systems,'' \emph{IEEE Trans. Veh. Technol.}, vol.~65, no.~1, pp.
  398--403, Jan. 2016.

\bibitem{1237406}
O.~Oyman, R.~Nabar, H.~Bolcskei, and A.~Paulraj, ``Characterizing the
  statistical properties of mutual information in {MIMO} channels,'' \emph{IEEE
  Trans. Signal Process.}, vol.~51, no.~11, pp. 2784--2795, Nov. 2003.

\bibitem{6503473}
O.~Ozel and S.~Ulukus, ``On the capacity region of the {Gaussian} {MAC} with
  batteryless energy harvesting transmitters,'' in \emph{2012 IEEE GLOBECOM},
  2012, pp. 2385--2390.

\end{thebibliography}

\end{document}